\documentclass{elsart}
\usepackage{graphicx,array}
\usepackage{epsfig}
\usepackage{cite}

\begin{document}

\begin{frontmatter}

\title{The Outer Tracker Detector of the HERA-B Experiment. \\
  Part III: Operation and Performance}

\large{HERA-B Outer Tracker Group}

\author[desyhh]{H. Albrecht},
\author[nikhef,utrecht]{T.S. Bauer},
\author[rostock]{M. Beck},
\author[desyz]{K. Berkhan},
\author[desyz]{G. Bohm},
\author[nikhef,utrecht]{M. Bruinsma},
\author[oslo]{T. Buran},
\author[desyhh]{M. Cape\'ans},
\author[tsinghua]{B.X. Chen},
\author[hub]{H. Deckers},
\author[ihep]{X. Dong},
\author[texas]{R. Eckmann},
\author[desyhh]{D. Emeliyanov},
\author[desyhh]{G. Evgrafov\thanksref{mephi}},
\author[dubna]{I. Golutvin},
\author[desyhh]{M. Hohlmann},
\author[desyhh]{K. H\"opfner},
\author[nikhef]{W. Hulsbergen},
\author[ihep]{Y. Jia},
\author[ihep]{C. Jiang},
\author[unido]{H. Kapitza\corauthref{cor}},
\corauth[cor]{Corresponding author. DESY, Notkestr.\ 85, D-22607 Hamburg,
  Germany. Tel.: +49\,8998\,3972; fax: +49\,8998\,4018. E-mail address:
  herbert.kapitza@desy.de (H. Kapitza).}
\author[rostock]{S. Karabekyan\thanksref{yerevan}},
\author[ihep]{Z. Ke},
\author[dubna]{Yu. Kiryushin},
\author[hub]{H. Kolanoski},
\author[hub]{D. Kr\"ucker},
\author[dubna]{A. Lanyov},
\author[tsinghua]{Y.Q. Liu},
\author[hub]{T. Lohse},
\author[hub]{R. Mankel},
\author[hub]{G. Medin},
\author[desyhh]{E. Michel},
\author[dubna]{A. Moshkin},
\author[tsinghua]{J. Ni},
\author[desyz]{S. Nowak},
\author[nikhef,utrecht]{M. Ouchrif},
\author[desyhh]{C. Padilla},
\author[rostock]{R. Pernack},
\author[desyhh,itep]{A. Petrukhin},
\author[dubna,hd]{D. Pose},
\author[desyhh]{B. Schmidt},
\author[desyz]{A. Schreiner},
\author[desyhh,rostock]{H. Schr\"oder},
\author[desyz]{U. Schwanke},
\author[desyhh]{A.S. Schwarz},
\author[desyhh]{I. Siccama},
\author[desyz]{K. Smirnov},
\author[dubna]{S. Solunin},
\author[desyhh]{S. Somov},
\author[desyz]{V. Souvorov},
\author[desyz,itep]{A. Spiridonov},
\author[desyz,hub]{C. Stegmann},
\author[nikhef]{O. Steinkamp},
\author[desyhh]{N. Tesch},
\author[desyhh]{I. Tsakov\thanksref{sofia}},
\author[hub,hd]{U. Uwer},
\author[dubna]{S. Vassiliev},
\author[dubna]{D. Vishnevsky},
\author[hub]{I. Vukoti\'c},
\author[desyz]{M. Walter},
\author[tsinghua]{J.J. Wang},
\author[tsinghua]{Y.M. Wang},
\author[desyhh]{R. Wurth},
\author[tsinghua]{J. Yang},
\author[ihep]{Z. Zheng},
\author[ihep]{Z. Zhu},
\author[rostock]{R. Zimmermann}

\address[nikhef]{NIKHEF, 1009 DB Amsterdam, The Netherlands\thanksref{nl}}
\address[texas]{Department of Physics, University of Texas at Austin,
  RLM 5.208, Austin TX 78712-1081, USA\thanksref{doe}}
\address[ihep]{Institute for High Energy Physics, Beijing 100039, P.R.\ China}
\address[tsinghua]{Institute of Engineering Physics, Tsinghua University, 
  Beijing 100084 , P.R.\ China}
\address[hub]{Institut f\"ur Physik, Humboldt-Universit\"at zu Berlin, 
  D-12489 Berlin, Germany\thanksref{bmbf}}
\address[unido]{Institut f\"ur Physik, Universit\"at Dortmund, D-44221 
  Dortmund, Germany\thanksref{bmbf}}
\address[dubna]{Joint Institute for Nuclear Research, 141980 Dubna, Russia}
\address[desyhh]{DESY, D-22603 Hamburg, Germany}
\address[hd]{Physikalisches Institut, Universit\"at Heidelberg, 
  D-69120 Heidelberg, Germany\thanksref{bmbf}}
\address[itep]{Institute of Theoretical and Experimental Physics, 
  117259 Moscow, Russia}
\address[oslo]{Dept.\ of Physics, University of Oslo, N-0316 Oslo,
  Norway\thanksref{no}}
\address[rostock]{Fachbereich Physik, Universit\"at Rostock, 
  D-18051 Rostock, Germany\thanksref{bmbf}}
\address[utrecht]{Universiteit Utrecht/NIKHEF, 3584 CB Utrecht, 
  The Netherlands\thanksref{nl}}
\address[desyz]{DESY, D-15738 Zeuthen, Germany}

\thanks[mephi]{Visitor from Moscow Physical Engineering Institute, 
  115409 Moscow, Russia.}
\thanks[yerevan]{Visitor from Yerevan Physics Institute, Yerevan, Armenia.}
\thanks[sofia]{Visitor from Institute for Nuclear Research, INRNE-BAS, 
  Sofia, Bulgaria.}
\thanks[nl]{Supported by the Foundation for Fundamental Research on
  Matter (FOM), 3502 GA Utrecht, The Netherlands.}
\thanks[doe]{Supported by the U.S.\ Department of Energy (DOE).}
\thanks[bmbf]{Supported by Bundesministerium f\"ur Bildung und Forschung, 
  Germany, under contract numbers 05-7BU35I, 05-7DO55P, 05-HB1KHA, 05-HB1HRA,
  05-HB9HRA, 05-7HD15I.}
\thanks[no]{Supported by the Norwegian Research Council.}

\date{\today}

\begin{abstract}
In this paper we describe the operation and performance of the HERA-B Outer
Tracker, a 112\,674 channel system of planar drift tube layers. The performance
of the HERA-B Outer Tracker system fullfilled all requirements for stable and
efficient operation in a hadronic environment, thus confirming the adequacy of
the honeycomb drift tube technology and of the front-end readout system.
The detector was stably operated with a gas gain of $3\cdot 10^4$ in an
Ar/CF$_4$/CO$_2$ (65:30:5) gas mixture, yielding a good efficiency for
triggering and track reconstruction, larger than 95\,\% for tracks with momenta
above 5\,GeV/c.  
The hit resolution of the drift cells was 300 to 320\,$\mu$m and the
relative momentum resolution  can be described as:
$\sigma(p)/p\,[\%] = (1.61\pm 0.02) + (0.0051 \pm 0.0006)\cdot p
\textrm{ [GeV/c]}$.
At the end of the HERA-B running no aging effects in the Outer Tracker cells
were observed. 
\end{abstract}


\end{frontmatter}

\section{Introduction}

HERA-B is a fixed target experiment at the 920\,GeV proton beam of the HERA 
electron-proton collider \cite{proposal,tdr}. Proton-nucleus interactions
are produced using an internal wire target in the halo of the proton beam.
The experiment was designed to study CP violation in $B$ meson systems,
requiring high interaction rates, good particle identification, and efficient
triggering.

\begin{figure}[htb] \centering
   \includegraphics[width=\linewidth,trim=30 230 0 0,clip]{%
     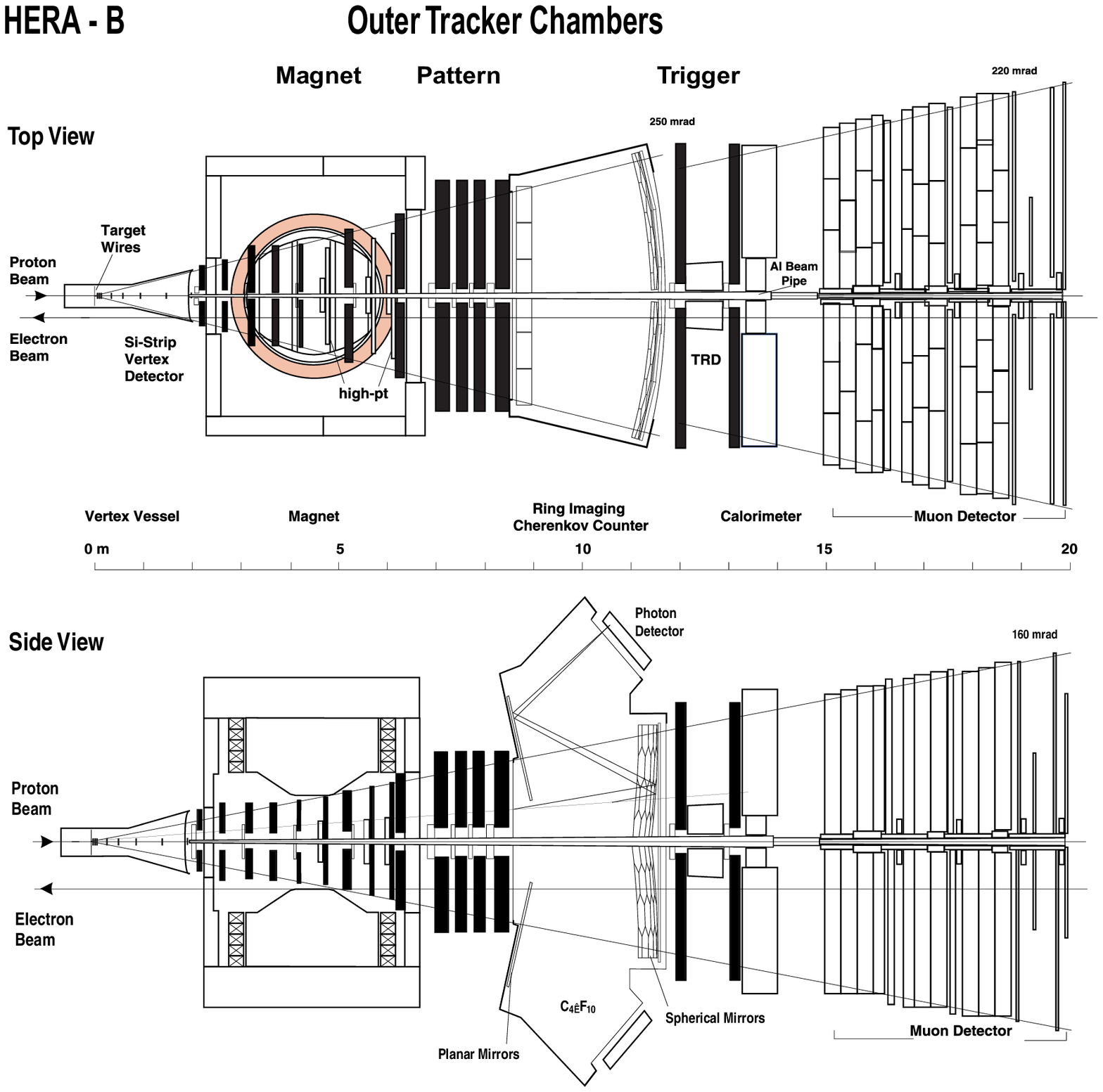}
   \begin{picture}(0,0)
      \put(-150,45){\includegraphics[scale=0.2]{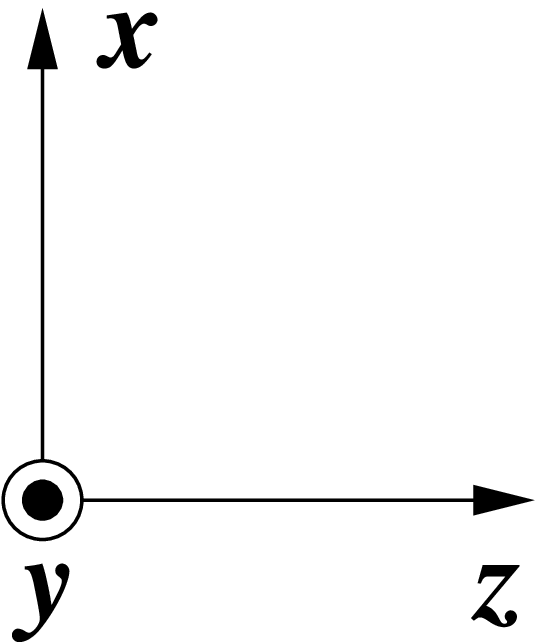}}
   \end{picture}
   \caption{\emph{Schematic top view of the HERA-B experiment with the Outer
       Tracker chambers indicated as black bars. The $z$-axis of the coordinate
       system coincides with the proton beam axis.}}
   \label{fig:herab}
\end{figure}

The main components of HERA-B are a silicon vertex detector (VDS), a
dipole magnet with a field integral of 2.1\,Tm, a tracking system consisting
of an Inner Tracker (ITR) using microstrip gas chambers and an Outer Tracker
(OTR) using drift tubes, and for particle identification a ring imaging
Cherenkov counter (RICH), an electromagnetic calorimeter (ECAL), and a muon
identification system (MUON) made of absorbers and drift chambers.
Figure \ref{fig:herab} shows a schematic top view of the experiment together
with the used coordinate system.

The Outer Tracker of HERA-B serves charged particle detection from the outer
acceptance limit of the experiment, defined by a 250\,mrad opening angle in
the bending plane of the magnet, down to a distance of 20\,cm from the HERA
proton beam. The OTR consists of 13 superlayers containing variable numbers of
planar honeycomb drift tube layers which provide three different stereo views
(wires at 0 and $\pm 80$\,mrad w.r.t.\ vertical). Seven superlayers are located
inside the magnet (labelled MC1--6,8), four between magnet and RICH (PC1--4),
and two between RICH and ECAL (TC1--2). Each superlayer consists of two
chambers which can be retracted from each other in $x$-direction. They are 
distinguished by
appending $+$ or $-$ to the superlayer label, depending on which side of the
proton beam tube the chamber is located. Around the proton beam the drift cell
diameter is 5\,mm, and the outer detector parts are equipped with 10\,mm 
cells. In
total 112\,674 electronic channels provide drift time information for tracking.
The detector is operated with the fast counting gas Ar/CF$_4$/CO$_2$ (65:30:5).
The details of the detector and of the electronics are described in the first
two papers of this series \cite{otr_design, otr_fee}. Pattern recognition and
track fit algorithms for the HERA-B tracking system are treated in
\cite{ranger}.

This paper is organized as follows: The next section describes the operating
conditions of the Outer Tracker. The long-term stability of several detector
components, including some results on chamber aging are also presented.
Its calibration and alignment, as well as the
achieved spatial resolution will be discussed in Section \ref{sec:calibration}.
Section \ref{sec:performance} summarizes the detector performance in terms of
hit and track efficiency and momentum resolution. Finally the paper is
summarized.

\section{Detector Operation}

\subsection{Running Periods and Conditions}
\label{sec:run}

The Outer Tracker was routinely operated during two data taking periods of
HERA-B which were separated by a long shutdown for the luminosity upgrade 
of HERA.

\subsubsection{Run Period 2000 (July 1999 -- August 2000)}

This run period started with an only partially installed OTR. 
After completion in January 2000, the up-time of the detector was 2200\,h. 
The first installed chambers had been operated for an additional 1100\,h.

By steering the movable wire target, the proton-target interaction rate was set
to values between 4 and 5\,MHz most of the time, resulting in maximum channel
occupancies of about 2\,\%.
For about 20\,h data
were also taken at high rates of 30 and 40\,MHz. Since the HERA bunch
separation is only 96\,ns, the high rate data contain multiple interactions
per event. In total the target produced about 33000\,MHz\,h corresponding to
$1.2\cdot 10^{14}$ proton-nucleus interactions in this run period.

Despite the stable performance of HERA and the good quality of the proton
beam, both resulting in very stable target rates, the OTR suffered from
severe HV stability problems during this run period (see Section
\ref{sec:HV_Stability}). In order to study these problems, the gas composition
and the HV settings were varied. As a result, not all data were taken 
with the OTR operating at its nominal gas gain of $3\cdot 10^4$.

\subsubsection{Run Period 2002 (December 2001 -- February 2003)}

For this run period the HERA-B physics programme focused on charm physics
\cite{rep2000}. A re-optimization of the detector for the new physics goal led
to the removal of the OTR superlayers MC2--8, with the aim to minimize the 
material in front of the ECAL.

Until the end of October 2002 priority was given to commissioning the upgraded
HERA accelerator and only 470\,h of beam time could occasionally be used by
HERA-B. In the following more continuous run the OTR was operated for another
760\,h.

Except for short tests, the detector was not operated at target rates beyond
5\,MHz during this run period. But instabilities of HERA operation, with spikes
of up to 2\,GHz in the proton-target interaction rate and strong positron beam
background,
caused large temporary charge depositions in the Outer Tracker chambers.
Occasionally currents of more than 50
times the nominal value were observed in the most exposed sectors of the
detector. Usually this triggered an automatic HV switch-off procedure. 
Providing one
chamber current as a feedback signal for beam steering to the HERA control room
strongly improved the running conditions for the Outer Tracker.
   
Due to the improved HV stability the operational parameters of the OTR could 
be kept stable during this run. However, during periods of high backgrounds
the detector performance suffered from distorted TDC spectra and reduced hit
efficiencies. Data from such periods were excluded from physics analyses and
performance studies. The useful data from
this run period include $210\cdot 10^6$ minimum bias and $148\cdot 10^6$
lepton-triggered events. Most performance results in this paper have been
obtained from these datasets.

\subsection{Slow Control System}
\label{sec:sloco}

During data taking most of the HERA-B hardware is monitored and controlled
by the Slow Control System \cite{slow_control}. From the Outer Tracker the
ASD-8 amplifier boards, TDC crates, temperature sensors, the high voltage
system, and the gas gain control are included.
The system allows to remotely read and set values
and to check the status of signals. The correctness of the controlled
parameters is monitored by the system. If a parameter value is outside a
predefined range,
an alarm message is generated. The high voltage control is described in
\cite{otr_fee}, the gas gain control is discussed in section
\ref{sec:stability_gain}. The OTR gas system \cite{gas_system} has its own
dedicated control hardware which is also connected to the detector safety
system. The system operates independently and only the gas parameters are
transmitted to the general HERA-B Slow Control system.

\subsubsection{ASD-8 Board Control}

Low voltage distribution boards are mounted on all superlayer frames. Each
board supplies up to 48 ASD-8 boards of similar sensitivity with power,
thresholds, and test pulses, and monitors voltages and currents. A CAN bus
connects the boards to the Slow Control System. More details can be found in
\cite{otr_fee}.

\begin{figure}[htb]\centering
   \includegraphics[width=0.8\textwidth]{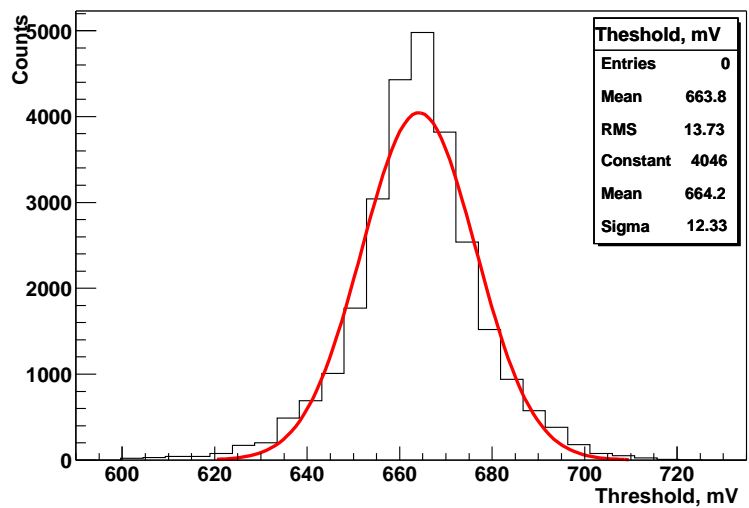}
   \caption{\emph{Threshold voltage distribution for a group of ASD-8 boards
       for the whole run period. A voltage of 660\,mV corresponds to a charge
       of 2.6\,fC.}}
   \label{fig:thresh}
\end{figure}

Threshold stability is an important prerequisite for reliable OTR resolutions
and efficiencies. The effective threshold voltage also depends on the ASD-8
supply voltages which are allowed to vary by about $\pm 10$\,\% ($\pm 300$\,mV)
before a
warning is issued. Figure~\ref{fig:thresh} shows the distribution of a
particular threshold voltage for the whole run period. The observed r.m.s.\ 
width of 14\,mV corresponds to a threshold charge uncertainty of about 0.1\,fC.

\subsubsection{TDC Crate Control}
\label{sec:craco}

Each of the crates mounted on the OTR superlayer frames contains TDC boards
and a daughter board of the Fast Control System (FCS) \cite{daq,fast_control}.
The crate control monitors the voltage and temperature of the power supply and
the speed of the cooling fan. It also allows to reset values of timing delays
and to control the crate power.

\subsection{Electronics Noise and Amplifier Thresholds}
\label{sec:enoise}

\begin{figure}[htb]\centering
   \includegraphics[width=0.9\textwidth]{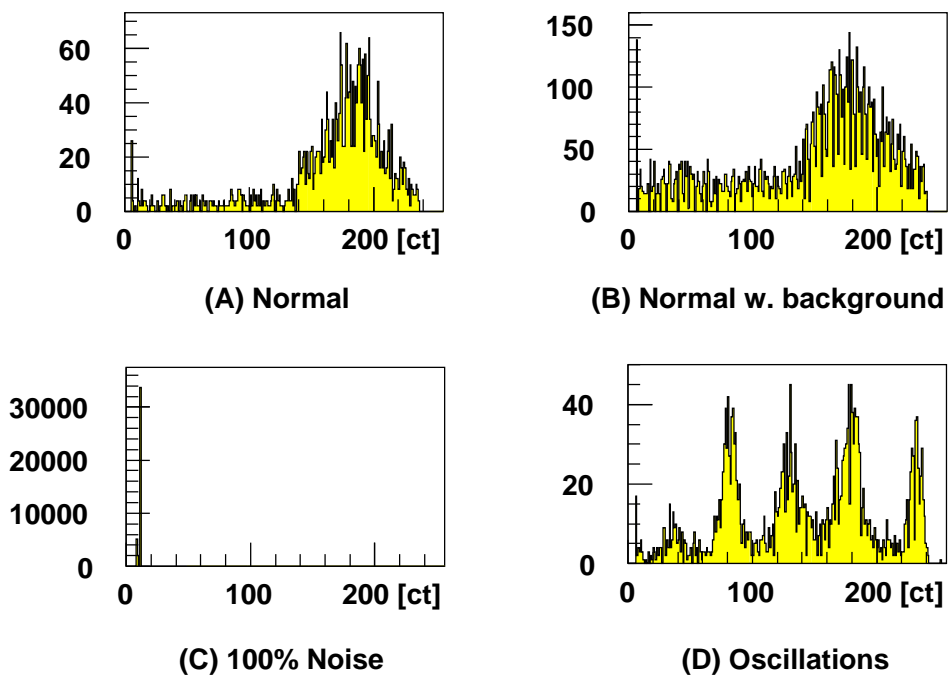}
   \caption{\emph{Typical examples of TDC spectra. A time range of 256 TDC
       counts corresponds to 100\,ns.}}
   \label{sp_types}
\end{figure}

The analysis of single channel TDC spectra is the key tool to monitor the
performance of every Outer Tracker channel. A normal spectrum is shown in
Fig.~\ref{sp_types} (A), while the other plots show the spectra of three 
classes of malfunctioning channels:
Channels with asynchronous noise (B),
channels with bad connections or shorts to ground (C), and
channels with oscillations (D).
For the data sample underlying the plots their fractions are 0.1\,\%,
1.8\,\%, and 8.8\,\%, respectively.

The asynchronous noise of type (B) is generated by other electrical devices
operating in the neighbourhood of the corresponding amplifier board. Because 
of different ground connections certain ASD-8 boards and channels are more 
affected than others. Since the timing of the electromagnetic noise is not 
correlated with the TDC clock, the time distribution of the noise hits is flat.
Improved ground connections between the gas boxes and their
ASD-8 boards almost completely eliminated this problem.

Most of the type (C) channels belonged to badly plugged ASD-8 boards. Such
faulty connections were regularly fixed during access days. Just a few of them
(0.014\,\%) were permanent due to a short to ground inside the gas box or on
the feedthrough board.

Bad ground connections of ASD-8 boards to the gas box and badly shielded
signal cables lead to oscillations of the amplifiers with a characteristic
frequency of about 50\,MHz which clearly shows up in Fig.~\ref{sp_types} (D).
This type of noise can only be suppressed by increasing the thresholds of the
corresponding ASD-8 boards.

For threshold specification in the experiment a parameter $\Delta U_{thr}$ is
used. It is added to the $U_{noise}$ value of each ASD-8 board.\footnote{%
  The ASD-8 boards for the Outer Tracker are sorted into 12 quality classes,
  using two characteristic threshold voltages: $U_{ref}$ which describes the
  response to a 4\,fC test pulse, and $U_{noise}$ which describes the noise
  susceptibility (see \cite{otr_fee} for the details).}
Using this parametrization one can use a common threshold parameter for all
boards of one chamber which usually come from several quality classes. This
makes it easy to react to changing noise conditions in the experiment. The
conversion factor between threshold voltages and charges is 250\,mV/fC.

In order to keep the thresholds $\Delta U_{thr}$ low and the loss of resolution
and hit efficiency as small as possible, the following measures were taken:
A direct connection of the analog ground of an ASD-8 board to the shielding of
its output cable, a shielding of the ASD-8 board input connector to reduce the
capacitive feedback from the digital output, a direct connection of the
digital and analog grounds right on the ASD-8 board, and modifications of the
TDC boards which reduced the feedback from their hit outputs. All these
measures reduced the number of noisy channels and allowed to set lower
thresholds on all chambers (see Table~\ref{thresh_noise}).
The performance results presented in the following are based on data from
the improved detector.

\begin{table}[htb]\centering
   \caption{\emph{Thresholds $\Delta U_{thr}$ of all OTR chambers after
       modification of the TDC boards. The old values are given in
       parentheses.}}
   \vspace*{2mm}
   \label{thresh_noise}
   \begin{tabular}{|cc|cc|}
      \hline\hline
      Chamber & $\Delta U_{thr}$ [mV] & Chamber & $\Delta U_{thr}$ [mV] \\
      \hline
      MC1-- & 100 (200) & MC1+ & 100 (200) \\
      PC1-- & 400 (600) & PC1+ & 400 (600) \\
      PC2-- & 500 (500) & PC2+ & 400 (500) \\
      PC3-- & 400 (500) & PC3+ & 400 (500) \\
      PC4-- & 400 (700) & PC4+ & 400 (600) \\
      TC1-- & 300 (600) & TC1+ & 300 (500) \\
      TC2-- & 300 (600) & TC2+ & 300 (500) \\
      \hline\hline
   \end{tabular}
\end{table}

\subsection{Gas Gain Control}
\label{sec:stability_gain}

Tests had shown that with the achievable thresholds for the
front\-end electronics a gas gain of about $3\cdot 10^4$ represents an optimum
in terms of efficiency, resolution, and operational safety of the Outer Tracker
chambers.
Some efforts have been taken to monitor and to control this important
parameter.

\subsubsection{Direct Gain Measurement}

A direct measurement of the gas gain has been performed at the beginning
of the 2002 run period. To determine the gas gain, the current in the OTR 
was measured as a function of the voltage applied to the chambers. In order to
reduce the uncertainty of the measurement at low voltages, where the current
approaches the ionization value, the total current from about 10000 drift 
cells (5\,mm) was measured. During the HV scan the interaction rate was kept
as stable as possible.

\begin{figure}[htb]\centering
   \includegraphics[scale=0.5,trim=30 90 0 0,clip]{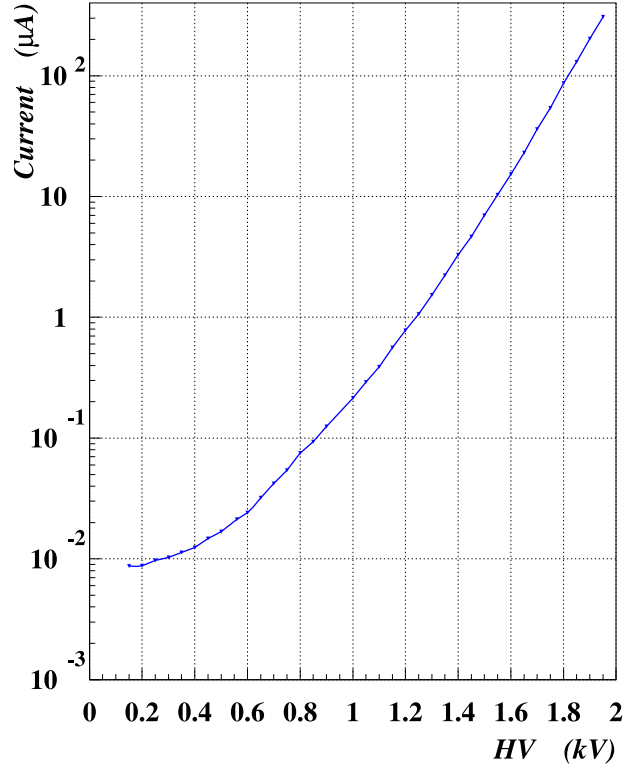}
   \caption{\emph{Sum current of $\approx 10^4$ drift cells (5\,mm) as a
       function of the voltage applied to the chambers.}}
   \label{current}
\end{figure}

The result is shown in Fig.~\ref{current}.
The gas gain $G$, defined as the ratio of the current determined at the
working point and the ionization current, is found to be
\begin{displaymath}
   G = \frac{I(1950\,\mathrm{V})}{I(150\,\mathrm{V})} = (3.3\pm 1.0)\cdot 10^4
\end{displaymath} 
This value agrees well with the results
of previous gain measurements using radioactive $^{55}$Fe sources.
The large error arises from uncontrollable fluctuations of the running
conditions among the points in Fig.~\ref{current} which were measured over a
period of several hours.

\subsubsection{Gas Gain Monitoring}

The amplification in the gas mixture strongly depends on ambient conditions 
like atmospheric pressure and temperature, as well as on the exact composition
of the counting gas. In particular the presence of oxygen has a large influence
on the gas gain.
For the Outer Tracker gas mixture Ar\,:\,CF$_{4}$\,:\,CO$_{2}$ the allowed 
limits for variation of the fractions are
($65\pm 1$\,:\,$30\pm 1$\,:\,$5\pm 0.2$)\,\%, which correspond to gain
variations of about 15\,\%. An Outer Tracker gas monitoring system was
installed to control these parameters and to act as an additional interlock if
they exceed the limits \cite{otr_design,gas_system}.

For gain monitoring a small chamber with 5 and 10\,mm honeycomb drift cells
of the OTR type is installed in parallel to the main gas input. The overlap
region of both cell types is irradiated by an $^{55}$Fe source. From the
measured pulse height spectrum the positions of the pedestal ($A_{ped}$) and 
of the 5.9\,keV gamma line ($A_{peak}$) are determined.
The gain value is given in \% with respect to a nominal one
\begin{displaymath}
   G = \frac{A_{peak}-A_{ped}}{A_{nom}}\cdot 100\,\%.
\end{displaymath}
$A_{nom}$ is the spectral position of the nominal gain.
The used technique allows to measure gains in the range (40--150)\,\%
with an accuracy of 3\,\%. The measurement takes up to 15 minutes.

In order to eliminate the large but trivial influence of the atmospheric
pressure on the gas gain, thus making smaller effects visible, a pressure
correction is introduced.
For small pressure variations this correction can be approximated by
\begin{displaymath}
   \frac{\Delta G}{G}=-K\frac{\Delta p}{p}.
\end{displaymath}
The corrected gain is obtained as
\begin{displaymath}
   G_{corr}=G[1+\alpha_{p}(p-p_{nom})],
\end{displaymath}
where $G$ and $p$ are the measured gain and pressure values,
$\alpha_p=K/p_{nom}=0.6\,\%$ per mbar is a pressure correction factor, and
$p_{nom}=1008$\,mbar is the nominal atmospheric pressure in Hamburg, averaged
over a year.

\begin{figure}[htb]\centering

   \resizebox{0.75\textwidth}{!}{

     \begin{picture}(450,640)

        \put(100,627){\scalebox{1.35}[1.35]{Gas Gain}}

        \put(196,628){\scalebox{1.1}[1.1]{measured}}
        \put(213,608){\scalebox{0.89}[1.1]{pressure corrected $-20\,\%$}}

        \put(196,625){\line(1,0){50}}
        \put(196,625){\vector(-1,-2){16}}

        \put(213,605){\line(1,0){78}}
        \put(291,605){\vector(2,-3){28}}

        \put(100,459){\scalebox{1.35}[1.35]{Atmospheric Pressure}}
        \put(100,292){\scalebox{1.35}[1.35]{Argon Content}}

        \put(82,258){\scalebox{1}[1.1]{Gas leak}}
        \put(82,255){\line(1,0){44}}
        \put(82,255){\vector(-2,3){14}}

        \put(100,123){\scalebox{1.35}[1.35]{Nitrogen and Oxygen Content}}

        \put(195,74){\scalebox{1.1}[1.1]{Nitrogen}}
        \put(195,71){\line(1,0){50}}
        \put(245,71){\vector(1,2){12}}

        \put(360,80){\scalebox{1.1}[1.1]{Oxygen}}
        \put(360,77){\line(1,0){43}}
        \put(360,77){\vector(-2,-3){12}}

        \put(82,83){\scalebox{1.1}[1.1]{Gas leak}}
        \put(82,80){\line(1,0){48}}
        \put(82,80){\vector(-2,3){14}}

        \put(242,491){\scalebox{1.2}[1.2]{Days}}
        \put(242,323){\scalebox{1.2}[1.2]{Days}}
        \put(242,155){\scalebox{1.2}[1.2]{Days}}
        \put(242,-12){\scalebox{1.2}[1.2]{Days}}

        \put(-5,570){\rotatebox{-90}{\scalebox{-1.2}[-1.2]{\%}}}
        \put(-14,410){\rotatebox{-90}{\scalebox{-1.2}[-1.2]{mbar}}}
        \put(1,240){\rotatebox{-90}{\scalebox{-1.2}[-1.2]{\%}}}
        \put(5,86){\rotatebox{-90}{\scalebox{-1.2}[-1.2]{ppm}}}

   \includegraphics[height=22cm,keepaspectratio]{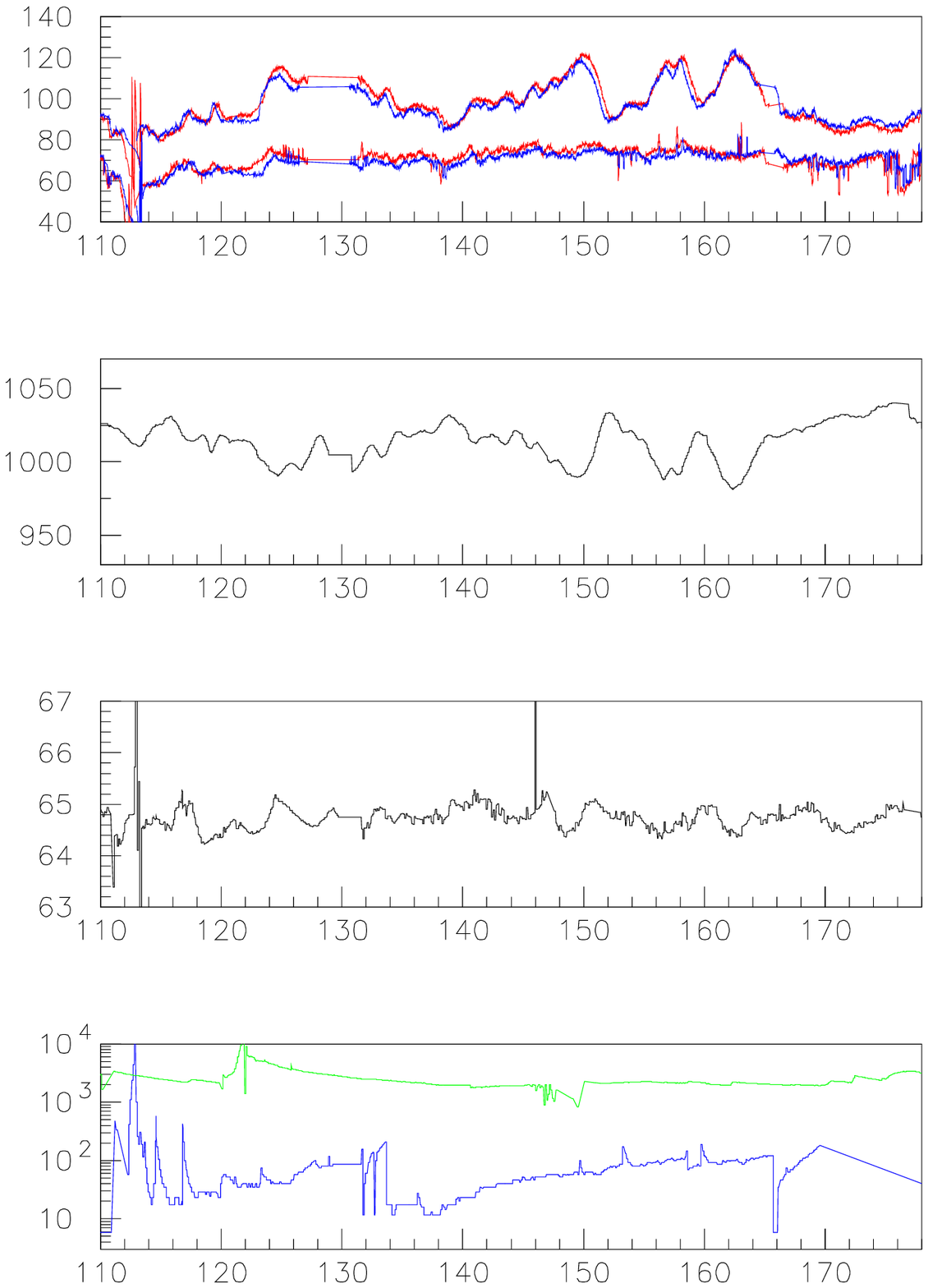}
     \end{picture}
     }
   \caption{\emph{Monitoring important gas system parameters. For clarity the
     pressure corrected gas gain curves for the 5 and 10\,mm cells are drawn
     with an offset of --20\,\%. Day 0 corresponds to August 20, 2002.}}
   \label{gain_2}
\end{figure}

The gas gain monitoring system is fully integrated in the
HERA-B Slow Control environment. Parameters like the atmospheric pressure, the
high voltage settings, the measured and the pressure corrected gains are stored
in a data base with a corresponding time stamp. The desired voltages for the
two Outer Tracker cell types, which follow from the high voltage dependence of
the gain, are also recorded. Figure~\ref{gain_2} illustrates the performance
of the system and shows that the largest gain variations are caused by the
varying atmospheric pressure. It also shows the stability of the mixing ratio
and the levels of O$_2$ and N$_2$ contamination.

\subsubsection{Online High Voltage Steering}

In order to operate the detector at an approximately constant gas gain, an
automatic adjustment of the high voltage is implemented in the OTR Slow
Control. The modified high voltage server retrieves the desired voltages from
the data base. A modification of the high voltage is considered to be necessary
if the currently applied voltages differ from the desired ones by at least 5 or
6\,V for the 5 or 10\,mm drift cells, respectively. Smaller differences are
not significant with the given gain measurement errors. The new high voltage
settings are not applied immediately but only when the Outer Tracker HV is
turned on next time.

For safety reasons the steering procedure is only used to balance gains in the
range ($70\ldots 140$)\,\%, corresponding to required voltage variations of
($+1.4\ldots -2.7$)\,\%. In addition, the desired voltage records from the
data base must not be older than one hour.

\subsection{Aging Investigations}

Aging of the Outer Tracker drift chambers was an extensively studied R\,\&\,D
subject \cite{herab_aging, etching} because of the high radiation load near the
beam pipes and since there was only little experience with the fast but
aggressive CF$_{4}$ gas component.


After exposure to a large radiation dose a wire chamber may exhibit an
irreversible gas gain reduction.
Since the chamber current at a given particle rate is proportional to the gas 
gain, the measured current can be used to monitor the gas gain.

\begin{figure}[htb]\centering
   \includegraphics[scale=0.5]{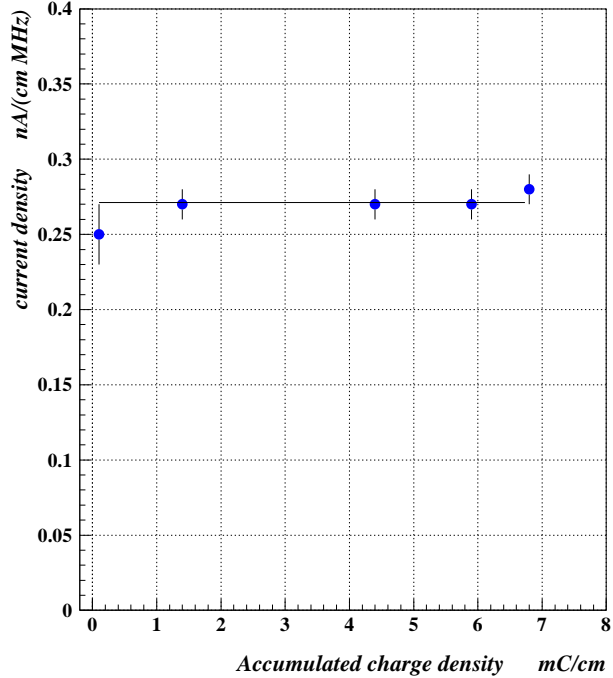}
   \caption{\emph{Current density as a function of accumulated charge.
       7\,mC/cm corresponds to 1000 hours of operation of HERA-B.}}
   \label{charge}
\end{figure}
The current density (current per cm of wire normalized to a target rate of
1\,MHz) as a function of the accumulated charge density is shown in 
Fig.~\ref{charge}. This dependence was measured in sector 4 of superlayer PC1 
which is situated close to the proton and electron beam pipes where the 
density of the accumulated charge is maximal (see \cite{otr_design} for sector
definitions).
As follows from  Fig.~\ref{charge}, there is no visible decrease of the 
current (and therefore of the gas gain) within an accuracy of 5\,\% for an
accumulated charge density up to about 7\,mC/cm.
This charge density corresponds to the operation of the HERA-B detector for 
more than 1000 hours during the 2002 run period. During its entire operation
time the Outer Tracker accumulated an average charge density of 24\,mC/cm on
the anode wires in the hottest regions.

Due to the much shorter than scheduled run of HERA-B and to mostly running at
target rates of 5\,MHz or even less, the accumulated charge density does not
seriously probe the OTR design criterion of being able to collect up to
1.5\,C/cm during five years of running at 40\,MHz.

After the run end an anode wire which was located near the proton beam pipe 
was taken out of the MC8 chamber and investigated using an electron microscope.
The wire surface is mostly clean, some very isolated depositions contain 
oxygen, silicon, chlorine, potassium, and sodium. There is no hint for wire 
etching which had been observed as a function of gas humidity in earlier
studies \cite{etching}.

\subsection{Detector Stability}
\label{sec:stability}

\subsubsection{High Voltage Stability}
\label{sec:HV_Stability}

The most disturbing OTR problem during the 2000 run was the continued loss of
high voltage groups. From the very beginning of OTR operation in July 1999 it
was observed that HV groups occasionally failed. The average long term rate of
these failures was one HV group per 5 hours up-time of the Outer Tracker, but
there were strong fluctuations. Since a HV group supplies 16 wires, the failure
of a single wire causes a channel loss which is a factor 16 higher.

\paragraph{Mechanical Instabilities:}
The failure rate was especially high in TC2--, the first assembled and
installed large chamber. Here it was suspected that this was caused by
mechanical instabilities of the modules, none of which was reinforced yet with
carbon fibre rods. Most modules in all later installed chambers had this
reinforcement \cite{otr_design}.
In a complete overhaul of TC2-- in December 1999 the chamber
was disassembled, all modules were reinforced and subjected to a more rigorous
HV test than before. With reference to a flat table, the module was bent with
a sagitta of 8\,mm, a condition similar to what was observed on modules
installed in TC2-- during the first assembly. All wires which then showed
shorts or excessive dark currents were disabled. Compared to the original test
procedure which only checked perfectly flat modules, the number of eliminated
wires roughly doubled, raising the fraction from 0.6 to 1.1\,\%. During the 8
months of running time after this repair only 16 of 606 HV groups in TC2--
showed problems, compared to 115 groups during the 5 months before the repair.

\paragraph{Shorts on HV Boards:}
By the end of the 2000 run there were 499 defect HV groups (out of 7219 in
total) in the whole Outer Tracker, not counting the TC2-- problems in 1999.
The inspection of damaged modules revealed a somewhat unexpected major source
of the problems: In more than 90\,\% of the cases the ``short'' was caused by
a burnt-in carbon trace of varying resistance across either of two capacitors
on the bottom side of the HV distribution board. In no case any of the 15 top
side capacitors caused the problem. This systematic difference had to do with
the differing mounting techniques used for top and bottom side components in
the SMD soldering process \cite{otr_fee}.
Less than 10\,\% of the faulty HV groups had genuine drift cell shorts which
were caused by mechanical module damage during installation, wire instabilities
in modules without reinforcement, and instantaneous wire breaking under
extremely high radiation load (rate spikes related to beam loss).

During the 10 months of the shutdown all chambers were disassembled and the
modules were subjected to the same repair steps as described above for TC2--.
The most time-consuming step in this procedure was the manual exchange of
12\,000 SMD filter capacitors on the HV distribution boards. 

\paragraph{Other losses of HV groups:}
Repaired and reinstalled chambers were kept at nominal high voltage for at
least three days, but the first repaired chambers were trained for as long as
82 days.
During this phase 55 HV groups were lost, mostly due to installation damage.
During the first phases of the 2002 run, when the OTR was only occasionally
operated, 33 HV groups were damaged. This was mostly related to
special events like a HV scan up to twice the nominal gas gain, or frequent
chamber movement for maintenance. For certain types of modules the internal
stability was not good enough to withstand the vibrations during chamber
handling. During the final 760\,h of regular OTR operation in 2002/2003 only
10 HV groups developed shorts, and more than half of these occurred during
extreme background conditions at the accelerator.

\subsubsection{Component Stability}

From 1999 to 2003 the complete OTR was operated for more than 3400\,h, some
chambers saw up to 1100\,h additional beam time (see Sec.~\ref{sec:run}). This
makes quantitative statements about the long term stability of various system 
components possible.

\paragraph{Detector Modules:}
The OTR was assembled from about 150 different types of detector modules
\cite{otr_design}. After reinforcement with carbon fibre rods the vast 
majority of these showed sufficient mechanical rigidity, resulting in good 
electrostatical stability during operation. In retrospect it must be said, 
however, that the modules with 5\,mm cells were designed too close to the 
stability limit, with free wire lengths of 59 and 68 cm between support strips
\cite{otr_design} in the PC and TC chambers, respectively. Most HV 
instabilities were observed in this region of the detector and 80 of 978 
modules were rebuilt when repairing the detector during the HERA shutdown.

\paragraph{Gas System:}
The most problematic components of the gas system were the metal bellow
circulation pumps \cite{gas_system}.
At times 5 of the 15 pumps were out of order thus reducing the gas flow rate
to only 70\,\% of the nominal value. In all cases the problems were caused by
leaking bellows. These leaks were most probably induced by solid contaminants
inside the closed gas loop. Occasional problems in the gas mixing system and
in the pressure regulation loops were mostly solved by recalibrating the mass
flow controllers. Only once such a problem was caused by the complete breakdown
of the steering unit.

\section{Detector Calibration}
\label{sec:calibration}

In order to obtain the spatial coordinates of the track hits from the measured
TDC values, several sets of calibration and alignment constants are necessary:
\begin{enumerate}
\item Bad channels are identified and recorded in channel masks which are
   used differently in online and offline tracking procedures.
\item The TDC spectra from all detector channels are corrected for individual
   time shifts $t_0$. These constants are generated by the
   $t_0$-calibration.
\item The translation of TDC counts to drift distances is done by means of a
   drift time to distance relation $r(t)$. The resolution
   function $\sigma(t)$ provides the errors of the measured distances. For
   each cell type both functions are generated by the $rt$-calibration.
\item Finally, the measured coordinates must be corrected by alignment
   constants which account for deviations of detector elements from their
   nominal geometry. The internal alignment determines Outer Tracker
   distortions, the global alignment the OTR displacement in
   HERA-B.
\end{enumerate}
A set of calibration constants remains valid as long as the corresponding
detector properties do not change. Failing hardware, threshold manipulation,
and chamber movements during the run periods call for several instances of
all sets. They are all stored in the HERA-B
Calibration-and-Alignment (CnA) Database \cite{CnA}.

\subsection{Masking of Bad Channels}

Bad Outer Tracker channels must be excluded from being used in pattern
recognition and track fit because they can disturb the procedures. They are
best identified in occupancy maps of the OTR chambers. An occupancy map is
a normalized hit map. For each wire it gives the number of recorded hits
divided by the number of analyzed events (channel occupancy). Bad channels are
found by comparing for every wire the measured channel occupancy with the 
occupancy predicted by a Monte Carlo (MC) simulation. Depending on whether 
their occupancies are much higher or much lower than expected, bad channels 
are called either noisy or dead.

Measured occupancies strongly depend on the interaction rate and on the trigger
configuration. In order to compare arbitrary experimental data with a standard
MC sample, measured channel occupancies are scaled by the ratio of the total 
numbers of hits in simulation and data. When this scaling is applied, there 
remains only a small influence of the trigger configuration, as was assessed 
by analyzing runs using different triggers. The influence of the material and 
the position of the target is also small.

In the MC simulation 300\,000 minimum bias events were generated and processed
through a detailed GEANT detector simulation \cite{geant}. This
is done only once per run period, assuming a stable geometry of
HERA-B. The calculated wire occupancies, the number of events, and the total
number of hits are stored.

For each set of experimental data 50\,000 events are analyzed. The measured
occupancies are properly scaled and compared to the MC reference. Noisy and
dead wires are identified using the cuts described below.
The generated channel mask is stored in the calibration database.

\begin{figure}[htb] \centering
  \includegraphics[scale=0.5]{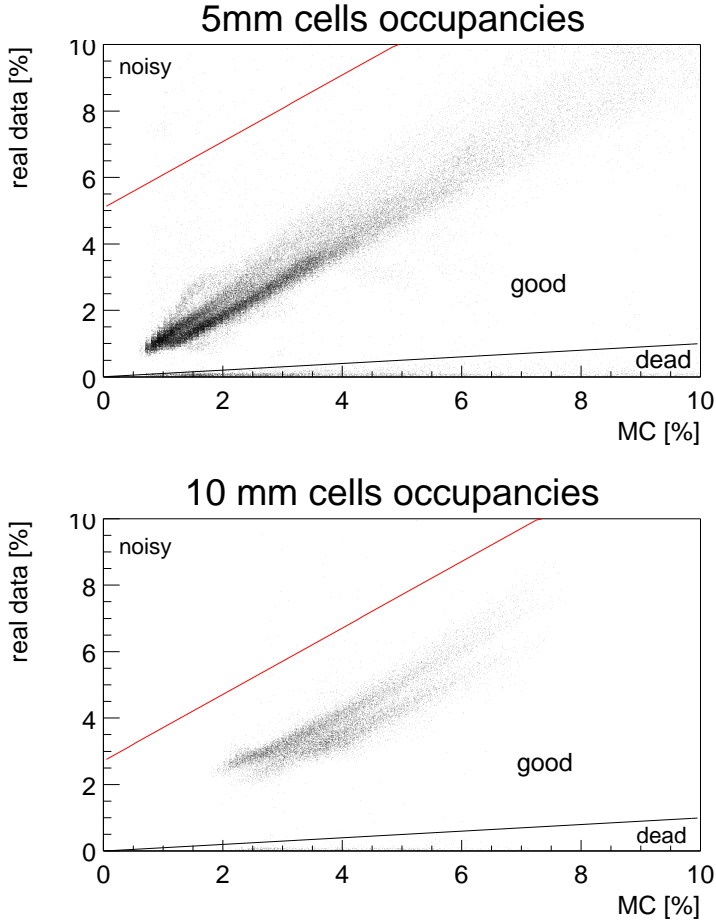}
  \caption{\emph{Distributions of measured vs. MC predicted occupancies for 
      5 and 10\,mm cells. Most noisy wires have occupancies close to 100\,\%
      and are out of range in this plot.}}
  \label{disertMasking2}
\end{figure}

An example for the distribution of measured versus MC predicted occupancies
for each wire is shown in Fig.~\ref{disertMasking2}.
Wires with measured occupancies $\Omega_{meas}$ smaller than 10\,\% of the MC
predicted value $\Omega_{MC}$ are marked as dead. This cut is indicated by the
lower lines in Fig.~\ref{disertMasking2}.
The optimal cuts to define noisy wires are
\begin{eqnarray*}
   \Omega_{cut} &=& \Omega_{MC} + \langle \Omega_{meas}\rangle\cdot 1.5\quad
                                                  \textrm{for 5\,mm cells,} \\
   \Omega_{cut} &=& \Omega_{MC} + \langle \Omega_{meas}\rangle\cdot 0.8\quad
                                                  \textrm{for 10\,mm cells.}
\end{eqnarray*}
They are indicated by the upper lines in Fig.~\ref{disertMasking2}. These
definitions accept a relatively large fraction of noise hits on wires for which
the simulation predicts small occupancies. This helps to maximize the number
of usable
channels and is tolerated by the track reconstruction algorithms as long as
the absolute number of hits remains low. More details and comparisons with
other masking methods can be found in \cite{ilija}.

Table \ref{NDGtable} shows for a typical run the fractions of good, noisy, and
dead wires in all chambers of the Outer Tracker.
\begin{table}[htb]\centering
   \caption{\emph{Fractions of good, noisy, and dead wires in all chambers of
      the Outer Tracker for a typical run.}}
   \vspace*{2mm}
   \begin{tabular}{|c|cc|cc|cc|}
      \hline\hline
      Superlayer & \multicolumn{2}{c|}{good  [\%]} 
                 & \multicolumn{2}{c|}{noisy [\%]}  
                 & \multicolumn{2}{c|}{dead  [\%]}\\ 
      side     & $+x$  & $-x$  & $+x$ & $-x$ & $+x$ & $-x$ \\ \hline
       MC1     & 93.1  & 94.5  & 0.1  & 0.1  & 6.8  & 5.4  \\
       PC1     & 93.1  & 94.7  & 0.4  & 0.5  & 6.5  & 4.8  \\
       PC2     & 94.2  & 94.1  & 1.2  & 0.6  & 4.6  & 5.2  \\
       PC3     & 94.1  & 94.8  & 3.0  & 1.1  & 3.0  & 4.2  \\
       PC4     & 91.7  & 95.9  & 0.8  & 0.3  & 7.5  & 3.8  \\
       TC1     & 90.4  & 93.2  & 0.5  & 0.3  & 9.2  & 6.5  \\
       TC2     & 91.2  & 90.8  & 0.7  & 3.0  & 8.1  & 6.2  \\
      \hline\hline
   \end{tabular}
   \label{NDGtable}
\end{table}

It should be noted that masked cells are treated differently in the different
parts of the trigger chain and in the track reconstruction. In order to prevent
an unacceptable loss of trigger efficiency, the First Level Trigger considers
noisy wires as good and dead cells as always having a hit, even though this
leads to an increased number of trigger messages. Contrary to this, masked
cells are excluded from the Second Level Trigger and from the online and
offline track reconstruction.

\subsection{TDC Spectrum Shifts}

The hardware calibration of the TDCs maps a time interval of 100\,ns linearly
to 256 TDC counts \cite{otr_fee}. Linearity and stability of the conversion 
factor are so good that it adds negligibly to the final uncertainty of the time
measurement. The origin of the time scale is not so well defined, the absolute
timing depends on the channel. In order to be able to use a common drift time 
scale throughout the Outer Tracker, the TDC spectra of all detector channels 
are time aligned using individual time shifts $t_0$. These calibration 
constants absorb systematic effects arising e.g.\ from differences in cell 
position or signal propagation time.

The $t_0$ calibration procedure evaluates the truncated means of the TDC
spectra \cite{t0}. The truncation limits $t_{min}$ and $t_{max}$, given in
Table~\ref{trunctiontable}, are chosen to accept 95\,\% of all hits.
The difference of the truncated mean and the reference value $t_{ref}$ yields
the time offset $t_0$ for each channel. The $t_{ref}$ values shown in
Table~\ref{trunctiontable} were chosen at the beginning of the run to
minimize the average $t_0$.
\begin{table}[htb]\centering
   \caption{\emph{Limits for the truncation of the TDC spectra and reference
       values for the $t_{0}$ calculation (all in TDC counts).}}
   \vspace*{2mm}
   \begin{tabular}{|c|ccc|}
      \hline\hline
      Cell Size & $t_{min}$ & $t_{max}$ & $t_{ref}$ \\ \hline
       5 mm     &       130 &       230 &      180  \\
      10 mm     &        50 &       250 &      150  \\
      \hline\hline
   \end{tabular}
   \label{trunctiontable}
\end{table}
Due to the truncation this extremely simple procedure must be iterated, but
already after the third iteration the $t_0$ values are stable within 0.5\,ns,
the time measurement bin of the TDC.

\begin{figure}[htb] \centering
 \includegraphics[width=\linewidth]{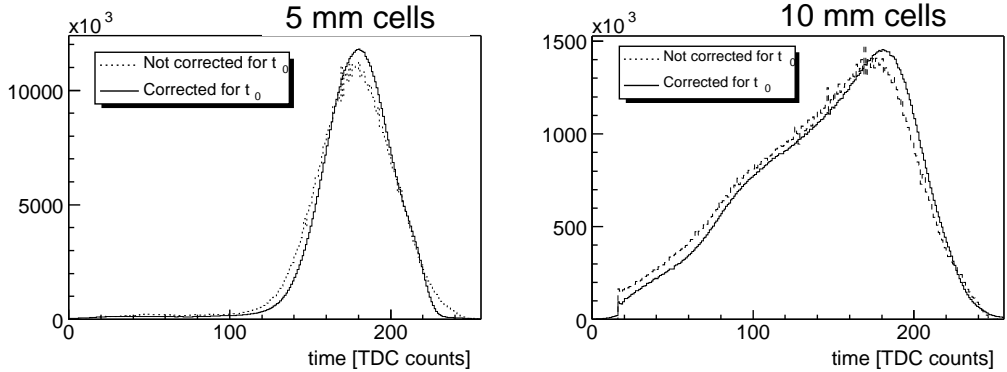}
 \caption{\emph{Summed TDC spectra of all cells (dotted line without, solid
     line with $t_0$ correction). Since the TDCs operate in ``Common Stop''
     mode, the time scale is inverted, i.e.\ large TDC counts correspond to
     short drift times and vice versa.}}
 \label{summedspectra}
\end{figure}

The quality of this $t_0$ calibration procedure can be assessed by comparing
summed TDC spectra before and after application of the $t_0$ corrections. As is
shown in Fig.~\ref{summedspectra} for both types of drift cells, the corrected
spectra have a higher peak and a steeper rising edge than the uncorrected ones,
just as one would expect from improving the overlap of the single wire spectra.
On the other hand, the effect is small which shows that the rough $t_0$ 
calibration done by setting delays on the Fast Control System (FCS) daughter
boards in the TDC crates is already quite good (see section \ref{sec:craco}).

\begin{figure}[htb] \centering
  \includegraphics[width=\linewidth]{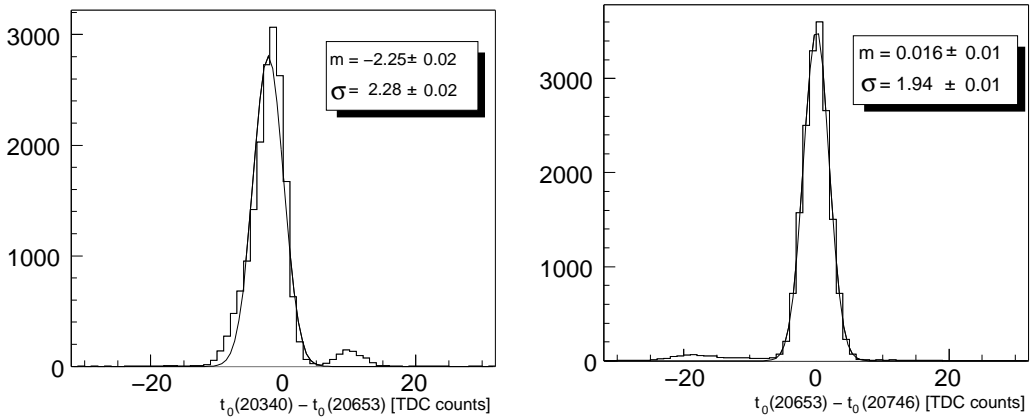}
  \caption{\emph{Long term stability of the $t_0$ calibration: Differences of
      $t_{0}$ values determined for the runs 20340, 20653, and 20746, taken
      during a time period of six weeks.}}
  \label{timestabilityoft0}
\end{figure}

The $t_0$ constants were determined for each run with more than 300\,000 
events.
As an illustration of the long term stability of the $t_0$ calibration, the
differences of the values obtained for three runs in a period of six weeks are
shown in Fig.~\ref{timestabilityoft0}. The shift of the left spectrum by
$-2.25$ TDC counts is a consequence of changed operational parameters.
The small structures at $+10$ and $-20$ TDC counts are attributed to
instabilities of the FCS delays for single crates. These were cured by
resetting the affected TDC crates.

\subsection{Drift Time to Distance Calibration}

The drift time to distance relation $r(t)$, or $rt$-relation for short,
transforms a measured drift time $t$ into a distance $r$ from the hit wire.
The resolution function $\sigma(t)$ provides the errors of the measured
distances $r(t)$. Both functions are needed when preparing hits for pattern
recognition and track fit. They are determined in a common $rt$-calibration
procedure. A detailed description of the HERA-B Outer Tracker $rt$-calibration
is given in \cite{t0}. Here only the main features of the procedure are
summarized and some results on the average hit resolution are presented.

\subsubsection{Calibration Procedure}

Using the drift distances $r(t_i)$ of all hits $i$ found by the pattern
recognition for a certain track, the track fit yields five parameters
which fully describe the track \cite{ranger}. Using these parameters, one can 
calculate distances of closest approach $d_i$ of the track to the anode wires 
of all used drift cells. While $r(t_i)$ is an unsigned quantity, namely the 
radius of a cylinder around the wire which the particle track must be tangent 
to, the calculated distance $d_i$ has a sign that tells on which side the track
has passed the wire. For a
quantitative comparison of measured and predicted distances the left-right
ambiguity of $r(t_i)$ is solved in the track fit by applying a sign parameter
$\epsilon_i$ such that the square of the residual
\begin{equation} \label{eq:residual}
   \Delta_i = \epsilon_i\,r(t_i)-d_i
\end{equation}
becomes minimal. After convergence of the track fit one has
$\epsilon_i=\mathrm{sign}(d_i)$ for every hit.

\begin{figure}[htb] \centering
  \includegraphics[width=0.49\linewidth]{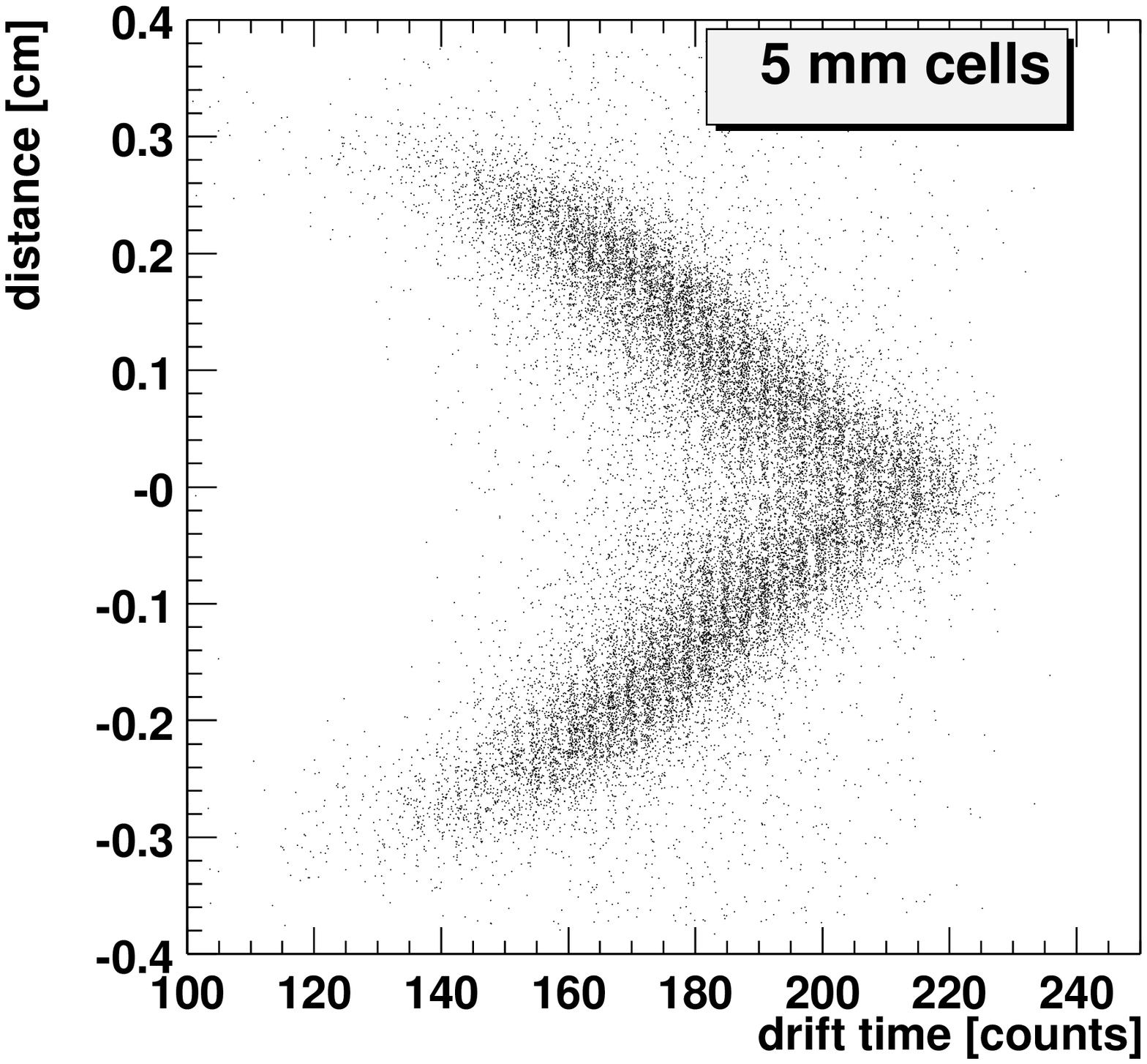}
  \includegraphics[width=0.49\linewidth]{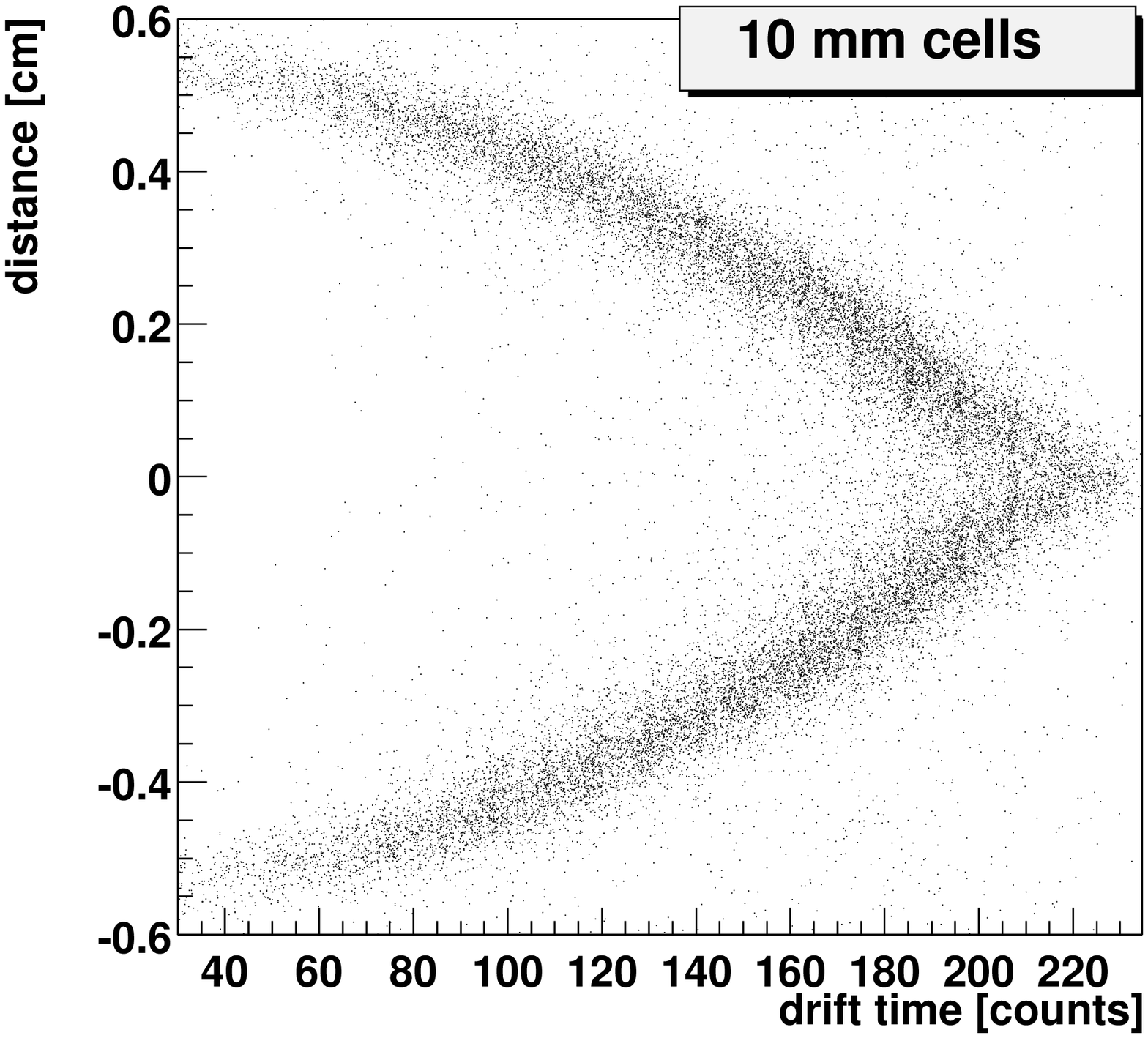}
  \caption{\emph{Signed reconstructed distance versus measured drift time for
      5 and 10\,mm cells in the PC chambers. One TDC count is 100/256\,ns.
      Because of operation in ``Common Stop'' mode high TDC counts correspond
      to short drift distances.}}
  \label{V_plot}
\end{figure}

The predicted distance $d_i$ depends on all hits participating in the 
track fit and therefore is a good estimator for the ``true'' distance of the 
track from the wire. Hence, if only an approximate $r(t)$ is used for the hit
preparation, an improved $rt$-relation can be obtained from the correlation of
predicted distances $d$ and measured drift times $t$. Figure~\ref{V_plot}
shows examples of this two-dimensional distribution for 5 and 10\,mm drift
cells in the PC superlayers.

\begin{figure}[htb] \centering
  \includegraphics[scale=0.4]{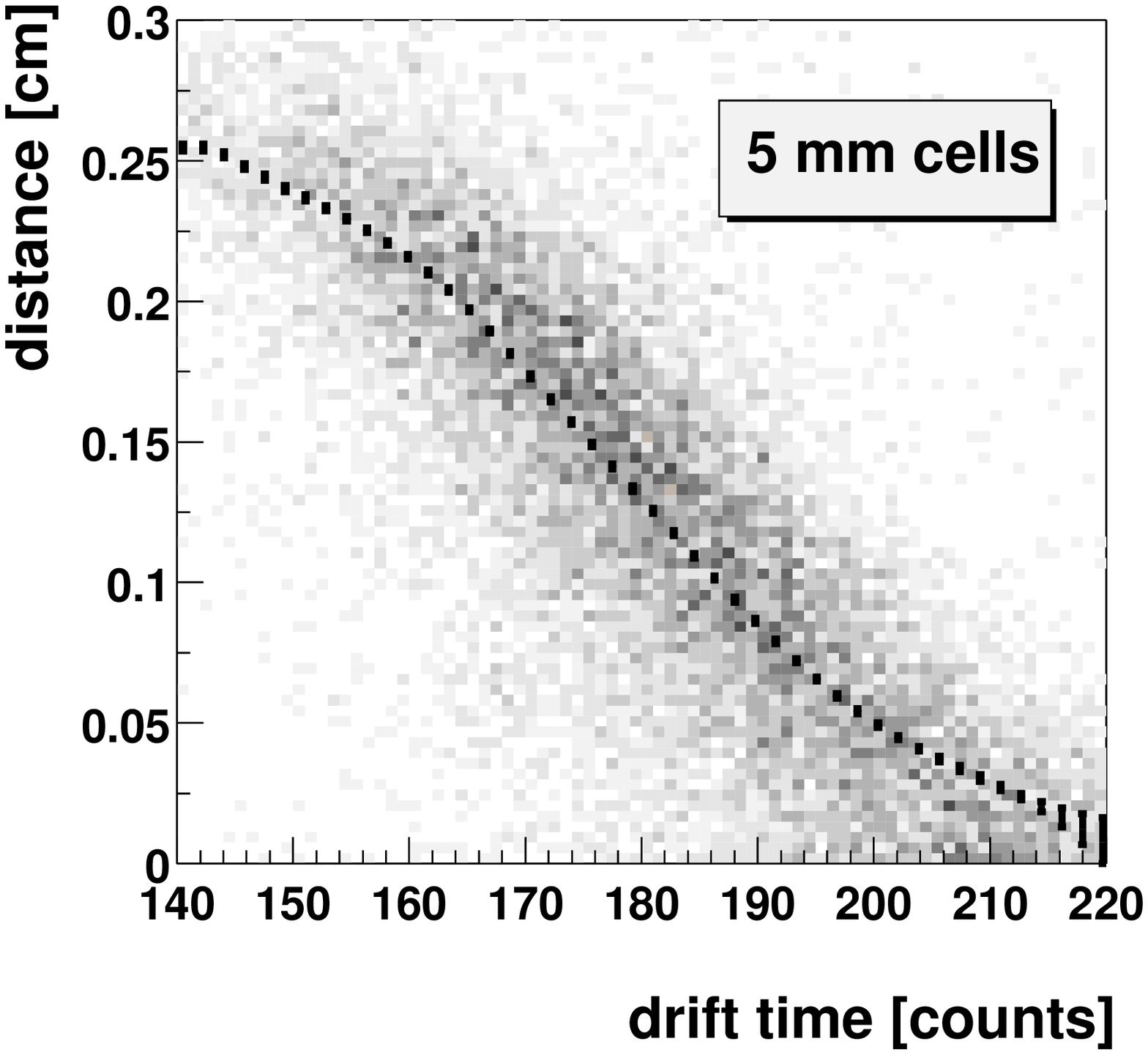}
  \includegraphics[scale=0.4]{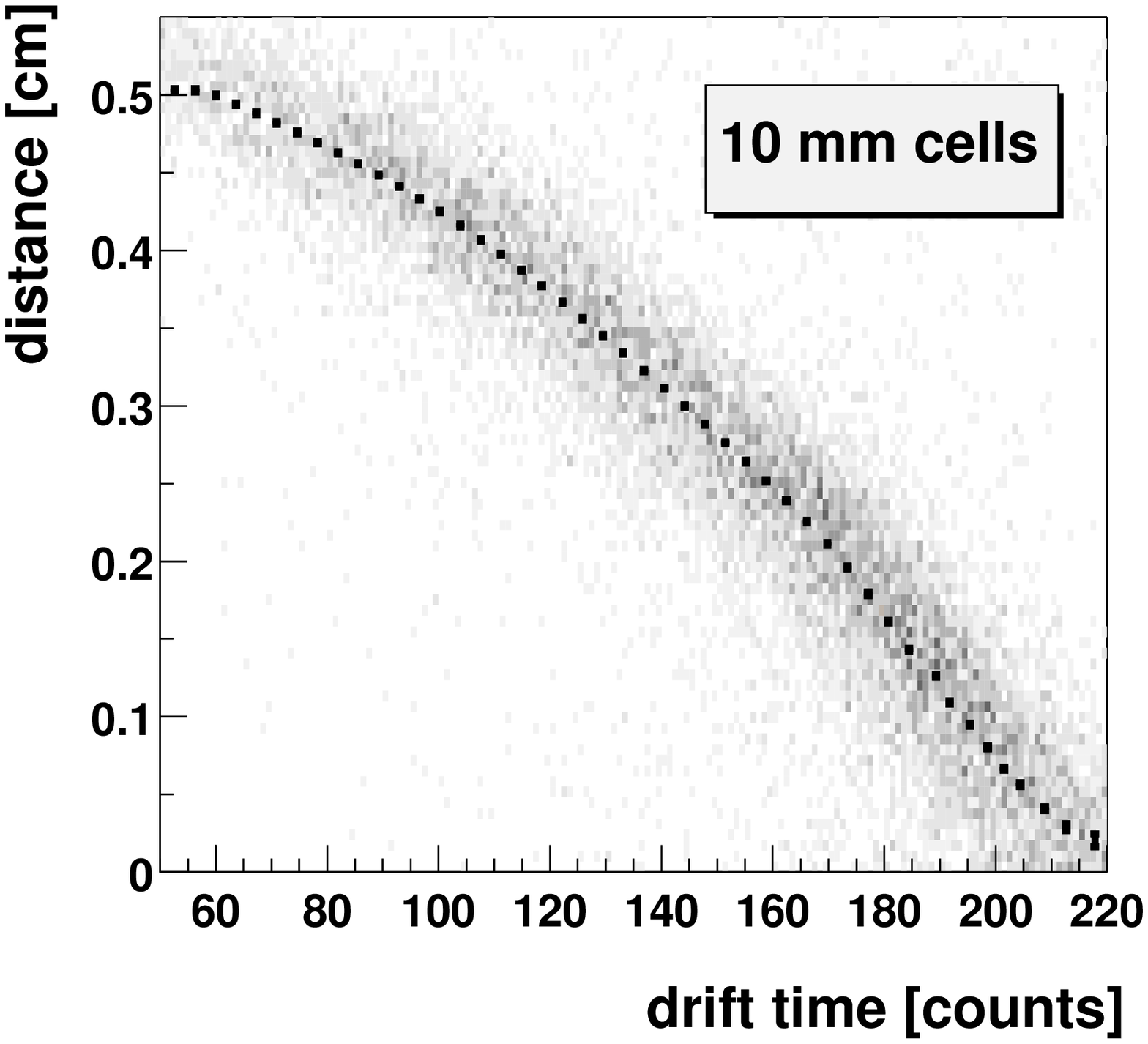}
  \caption{\emph{Interpolated points of the $r(t)$ functions for 5 and 10 mm
      cells. The input distributions $|d|(t)$ are shown as well.}}
  \label{rt_relat_rez}
\end{figure}

In each drift time bin the average value $\langle|d|\rangle$ of the unsigned
predicted distances is calculated. The resulting data points are smoothed in a
least squares fit using third order B-spline polynomials. These eventually
yield the values of the new $rt$-relation.
Figure \ref{rt_relat_rez} illustrates such an $rt$-calibration step. From the
plots one can also deduce quite high drift velocities of about 115 and
90\,$\mu$m/ns in the centers of the 5 and 10\,mm drift cells, respectively.

\begin{figure}[htb]
  \includegraphics[width=0.49\linewidth]{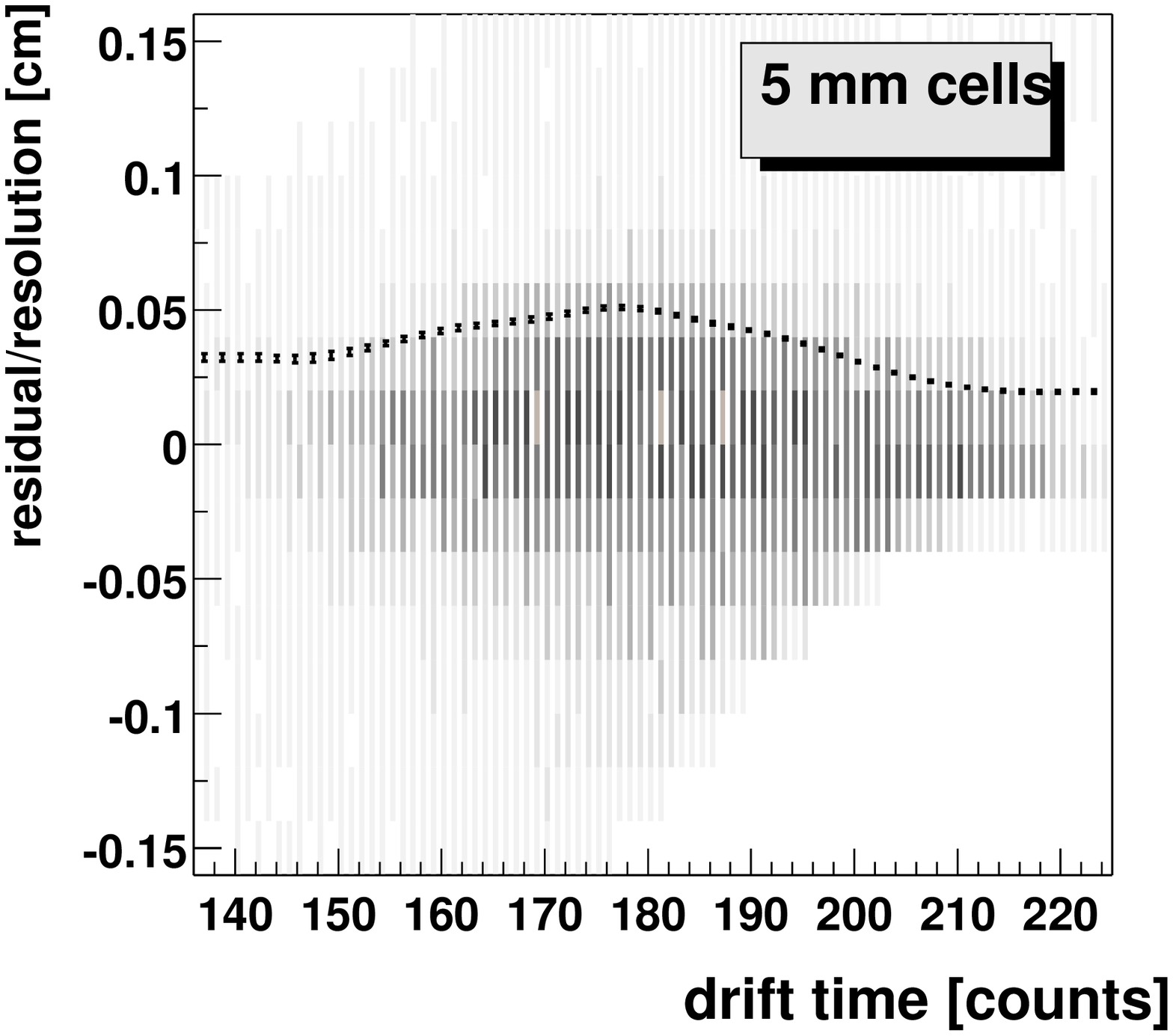}
  \includegraphics[width=0.49\linewidth]{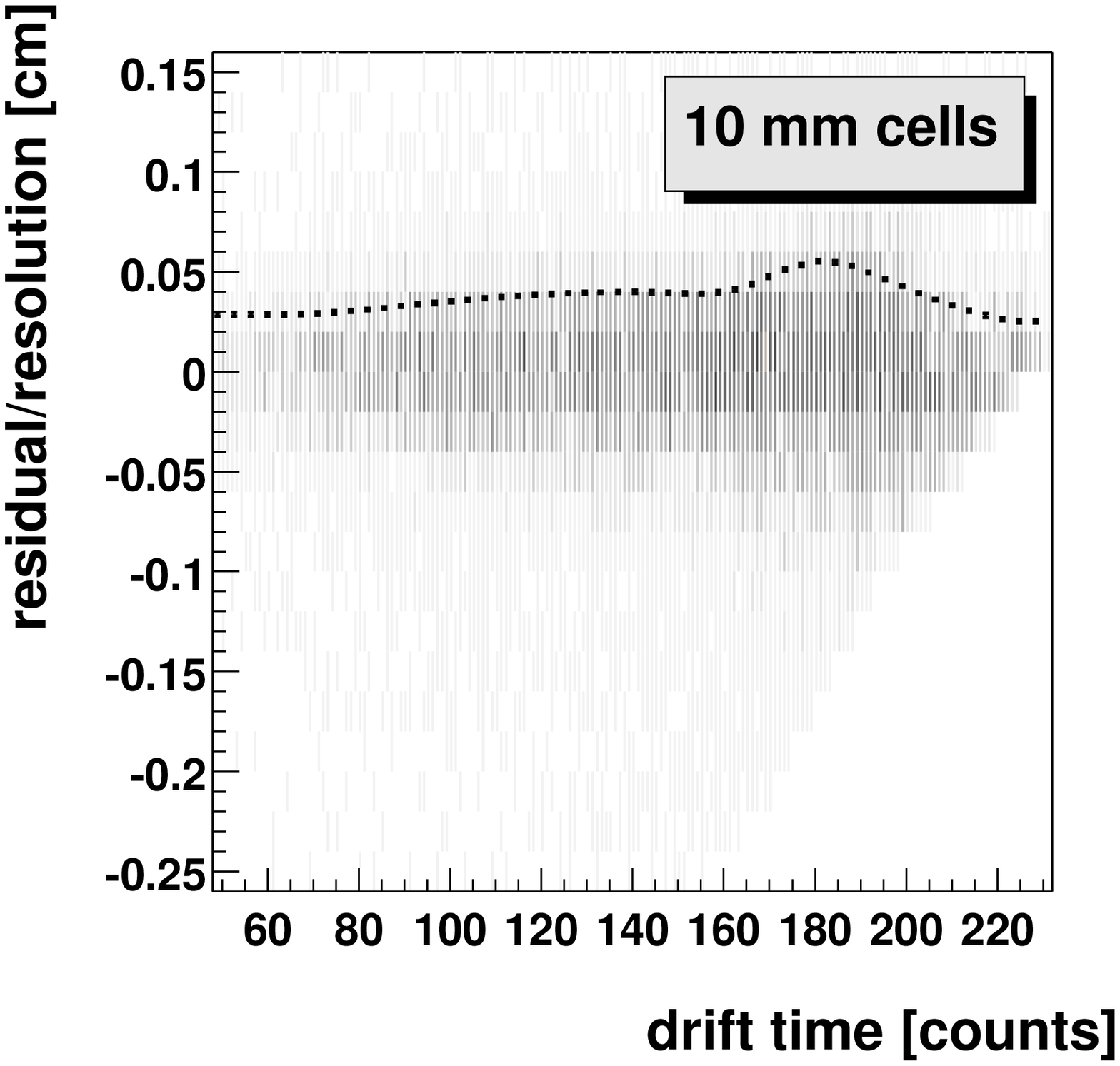}
  \caption{\emph{Resolution functions $\sigma(t)$ for 5 and 10\,mm drift cells
      (dotted lines). The input distributions $\Delta(t)$ are shown as well.}}
  \label{rezid_relat_rez}
\end{figure}

The value of the resolution function $\sigma(t)$ in a given drift time bin
follows from the RMS width of the distribution of residuals (\ref{eq:residual})
in that time bin. Also here the measured points are smoothed using a B-spline
fit. Since all hits which populate the residual distributions are also used
in the track fit, an individual scaling factor must be applied to each
residual, in order to compensate for the introduced bias \cite{t0}.
Figure \ref{rezid_relat_rez} shows distributions of scaled residuals as a
function of the measured drift times for both types of drift cells. For each
time bin the RMS width of the residual distribution, i.e.\ the value of the
new resolution function is shown as a point.

The described steps of obtaining new functions $r(t)$ and $\sigma(t)$ define
an iterative procedure. Typically five iterations are sufficient to converge
to the required precision. For track reconstruction in the first iteration 
$r(t)$ is obtained by integrating the drift time spectrum and the resolution is
assumed to be constant at 1\,mm.

The $rt$-relation discussed here is a tabulated function which is used for the
transformation of TDC counts to distances. It is determined using tracks going
through the whole detector and thus is influenced by global effects like
variations between drift cells, misalignment, variations in interaction time,
and by properties of the track reconstruction procedure itself. In certain
details this phenomenological function may be quite different from the physical
$rt$-relation of a drift cell which is defined by the operating parameters of
the cell (e.g.\ electric and magnetic field, drift gas mixture, pressure,
temperature) and modified by properties of the attached readout electronics
(e.g.\ impedance, threshold, pulse shaping).

The same arguments apply for the resolution function. The improvement of
$\sigma(t)$ towards the sense wire (large TDC counts in Fig.\
\ref{rezid_relat_rez}) which contradicts the expectation from drift chamber
physics, results from the described method of resolving the left-right
ambiguity. This method suppresses negative residuals $\Delta(t)<-r(t)$ and
leads to asymmetric and artificially narrow distributions close to the wire.
Here $\sigma(t)$ still yields the correct hit weight for the track fit, but it
is no longer identical to the physical drift chamber resolution.

The technical nature of both functions is emphasized by using a hardware
defined time scale of TDC counts instead of proper drift times in nanoseconds.

\subsubsection{Performance on Data}

\begin{figure}[htb]
   \includegraphics[scale=0.3]{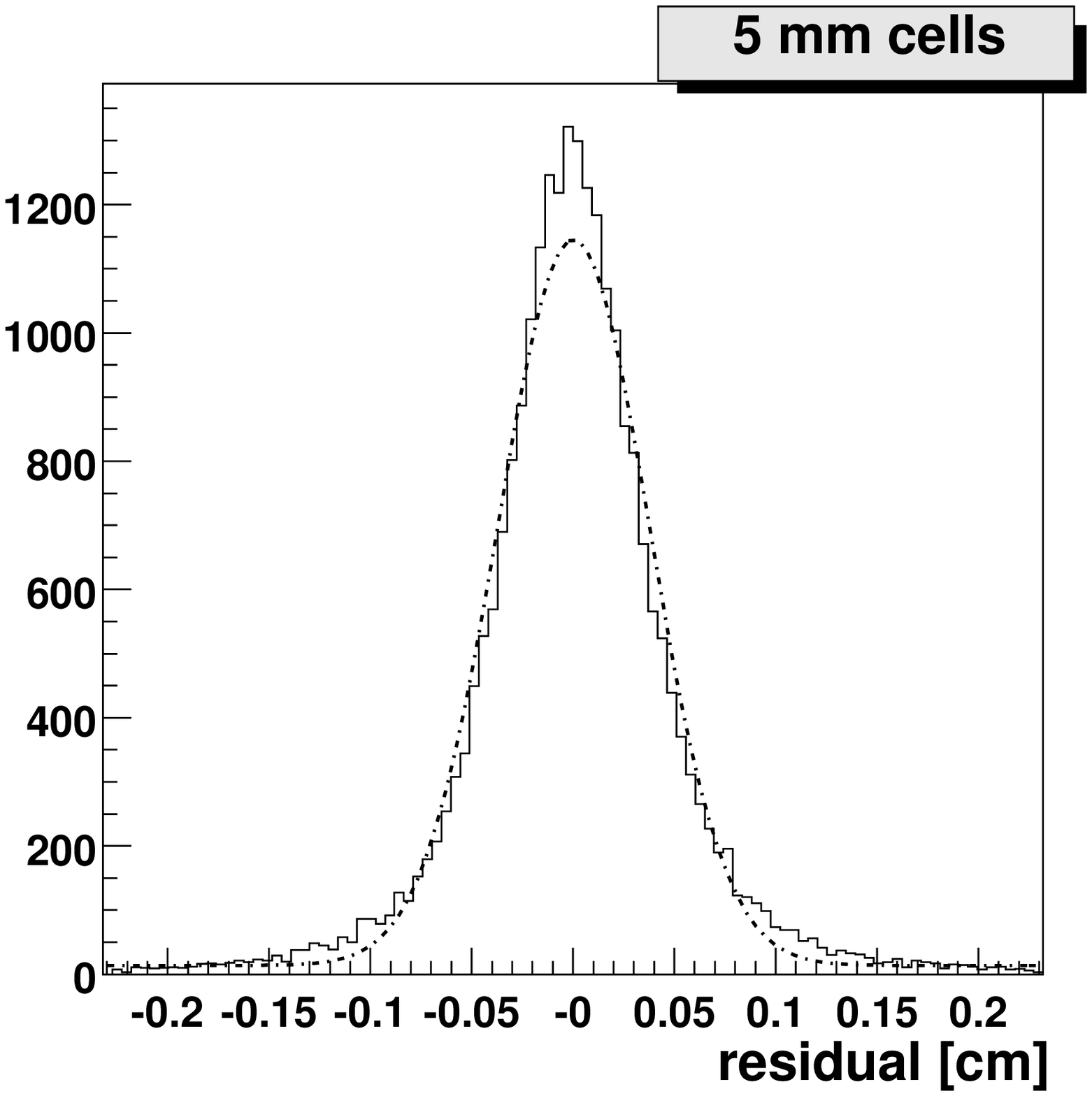}
   \includegraphics[scale=0.3]{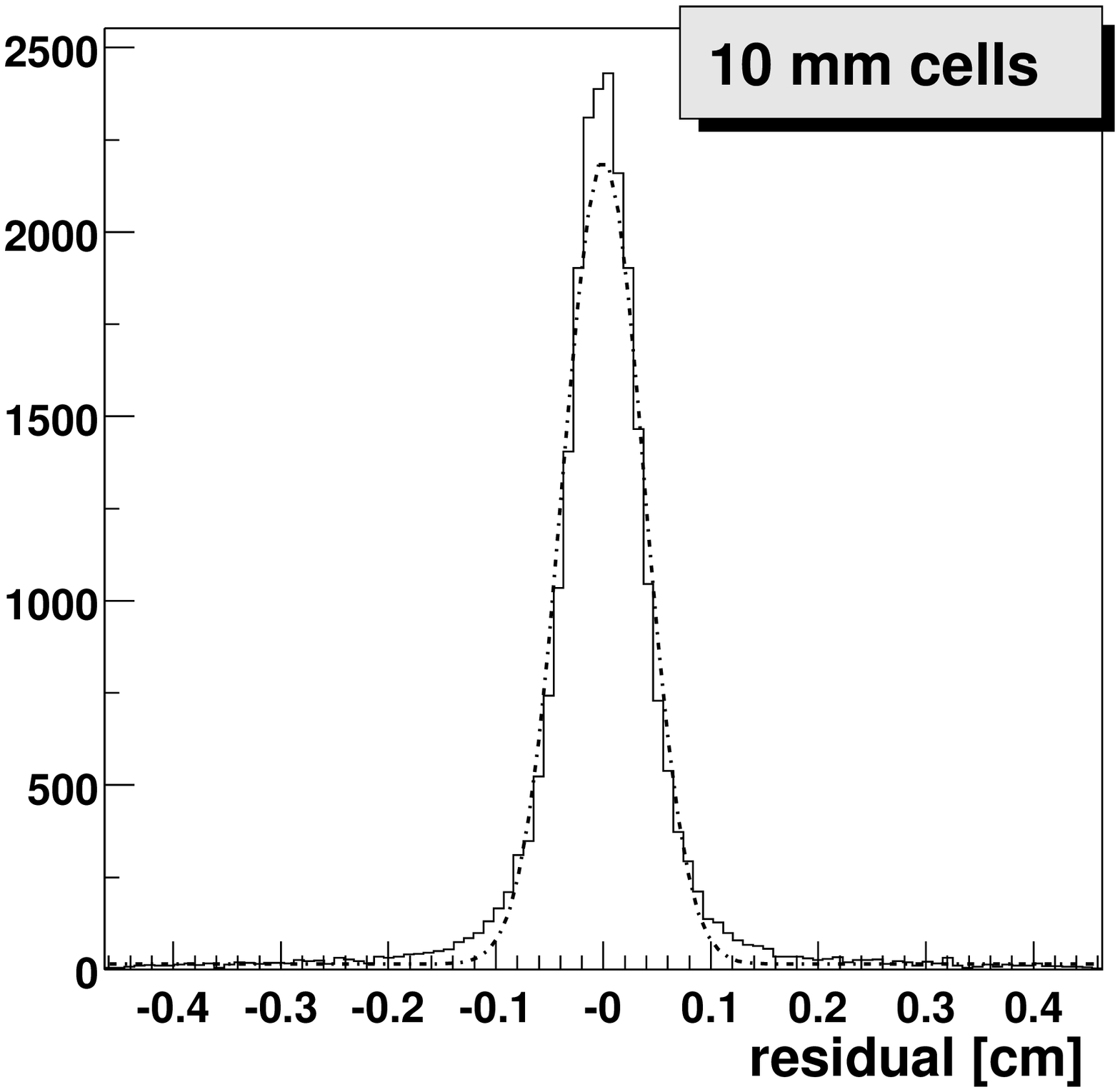}
   \caption{\emph{Distributions of scaled residuals for 5 and 10 mm cells in
       the PC chambers of the Outer Tracker. The dashed lines represent
       Gaussian fits. The average resolution is about 370\,$\mu$m in both cell
       types.}}
   \label{rez_5mm_10mm}
\end{figure}

Figure \ref{rez_5mm_10mm} shows the distributions of scaled residuals for 5
and 10 mm cells in the PC chambers of the Outer Tracker. In both types of
cells an average resolution of about 370\,$\mu$m is obtained. This includes a
substantial contribution from multiple scattering because the chambers in the
PC region represent an average thickness of 0.15 radiation lengths while for
statistics reasons the calibration procedure uses a track momentum cut of only
5\,GeV/c.

\begin{figure}[htb] \centering
   \includegraphics[width=\linewidth,trim=0 0 0 20,clip]{%
       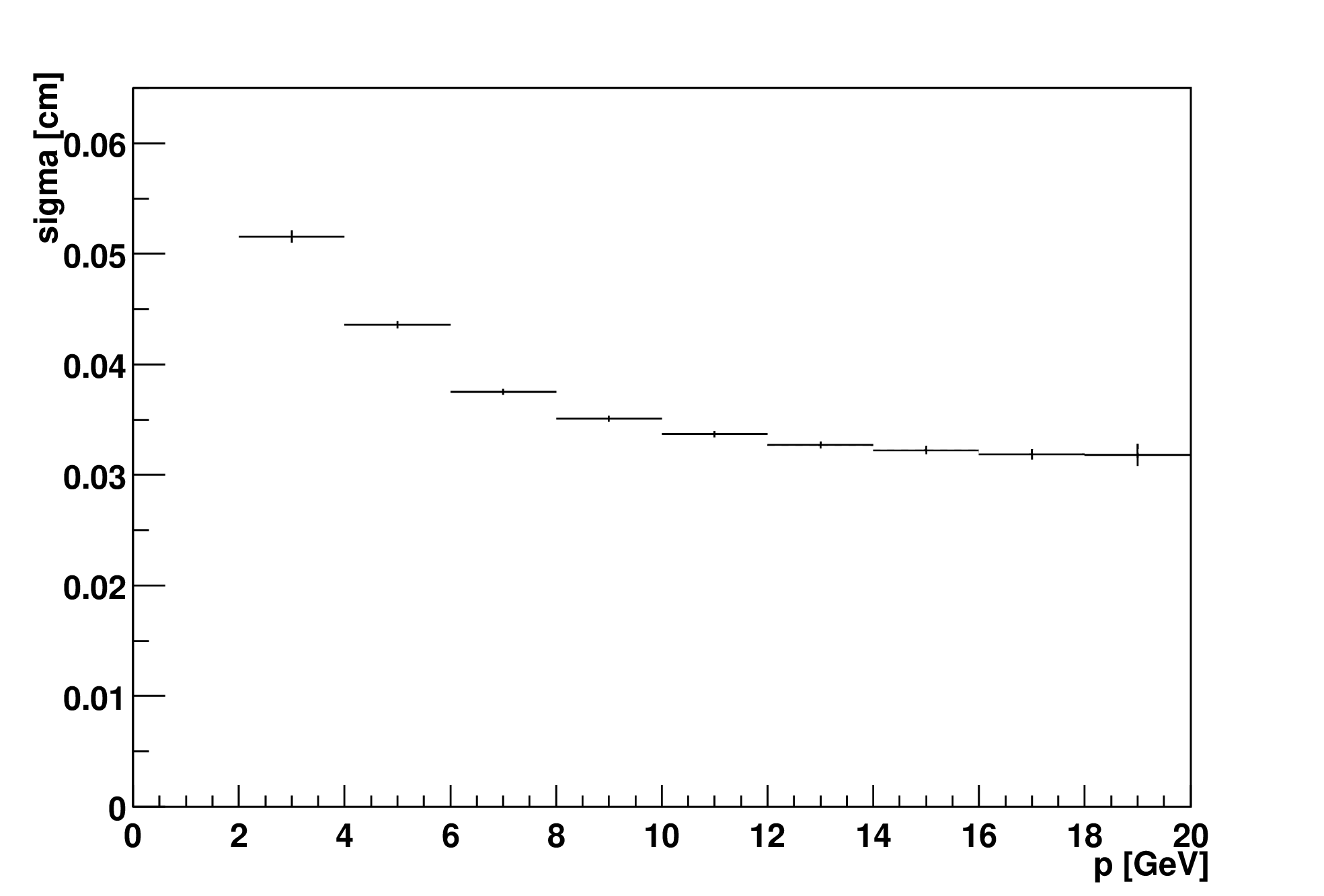}
   \caption{\emph{R.m.s. width of the residual distribution for 5\,mm drift
       cells as a function of track momentum.}}
   \label{res} 
\end{figure}

In order to investigate this further, the average resolution of the 5\,mm cells
is determined as a function of track momentum.
For this purpose high quality tracks are selected which have at least 20 hits
in the PC chambers and which are also observed in the silicon vertex
detector and in the TC chambers. In each of the chambers PC2 and PC3 one
layer is excluded from pattern recognition and track fit. In these layers
distributions of unbiased residuals are accumulated in bins of track momentum.
The result is shown in Fig.~\ref{res}.
The measured data are compatible with a chamber resolution of 300\,$\mu$m and
multiple scattering contributions of 80 and 310\,$\mu$m at $p=20$ and 5\,GeV/c,
respectively.

A chamber resolution of 300\,$\mu$m falls outside the range 150--250\,$\mu$m,
the design target in the proposal \cite{proposal}, which was verified in
several series of test measurements \cite{TDCtimestability}. This is mostly
due to the elevated noise level in the full detector compared to limited test
setups (see Sec.~\ref{sec:enoise}).
The required higher settings for the discriminator thresholds worsen
the resolution.
For a large system with more than 100\,000 channels also the threshold
granularity becomes important \cite{otr_fee} because it defines how closely
the detector can be adapted to spatially varying noise conditions.
Finally, a full tracking
system of independent chambers also suffers more from alignment and calibration
uncertainties both of which adversely affect the spatial resolution.

\subsection{Alignment}

There are two independent alignment tasks:
The internal alignment deals with the relative positions of the
individual OTR modules, while the global alignment determines the
spatial position of the OTR with respect to the other components of HERA-B.

\subsubsection{Internal Alignment Considerations}
\label{internal_alignment}

An internal alignment procedure for the HERA-B Outer Tracker must meet certain
demands:
\begin{itemize}
\item It must use tracks reconstructed in the OTR itself because there are no
   external tracks of sufficient precision available.
\item It must be able to work with hadronic tracks that are found in typical
   HERA-B events with only a moderate momentum cut being applied (typically 
   5\,GeV/c). Such a track sample can be collected within a few hours and 
   covers the full acceptance of the detector. This is adequate for the 
   sometimes quite short intervals of stable detector geometry. 
\item It must be robust against a possible bias introduced by the track 
   reconstruction procedure. The problem with the above track sample is the
   large spatial track density per event and hence the proper assignment of
   hits to tracks. If no measures are taken any track reconstruction algorithm
   will pick up the nearest hits and will prefer left-right signs for the
   drift distances that minimize the residual squares, thus creating a
   background in the residual distribution that peaks at zero.
\end{itemize}
In the following section an iterative algorithm is described which meets the 
above demands. Possible improvements by other algorithms have been investigated
\cite{belotelov} but have not become relevant for the HERA-B data analysis.

\subsubsection{Iterative Method of Internal Alignment}

The HERA-B geometry is described by a data structure called GEDE (GEometry of
DEtector). In the following this name will be used to denote the smallest OTR
units to be aligned, the sensitive sectors of OTR modules \cite{otr_design}.
There are in total 1380 GEDE units within the PC/TC area.
GEDE units belonging to the same module are mechanically coupled but
for simplicity such constraints are neglected. Each GEDE unit $k$ is
treated as a rigid body and two 
alignment constants are considered:
$a^u_k, a^z_k$, 
where $u$ means the direction orthogonal to
the wire, $z$ is the direction along the beam. The stereo angle $\alpha$ is 
treated as a constant.

If the alignment is based on internal tracks, all GEDE parameters become
correlated. Instead of a proper treatment of the large correlated system
we simplify the problem and consider each GEDE as an independent unit. 
The algorithm is then based on the residual distributions for each
individual GEDE plane. To avoid the above mentioned uncertainties of the
track-to-hits association, for each track the residuals are calculated
for all hits in the traversed GEDE.
Also for the left-right ambiguity of the drift distance measurement no decision
is made at this point, but both signs are accepted. This is done by creating
two residual histograms $h_+$ and $h_-$ per GEDE plane, which are filled
using always the positively or negatively signed drift distance, respectively.
The left hand plots in Fig.~\ref{histexample} show the characteristic shapes of
these histograms:
\begin{figure}[htb] \centering
   \includegraphics[width=\linewidth]{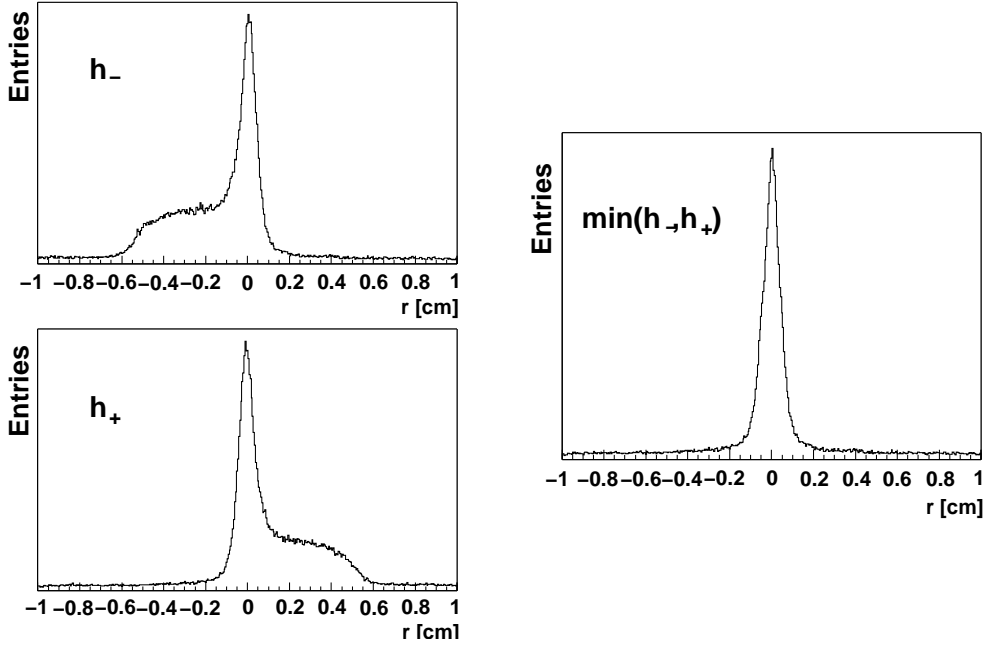}
   \caption{\emph{Combining two residual histograms using always positive or
       negative drift distance signs.}}
   \label{histexample}
\end{figure}
There is a flat background from calculating residuals using hits in wrong
cells, a shoulder from calculating residuals using hits from the right cell
but with wrongly signed drift distances, and a peak resulting from properly
calculated residuals. The peak occurs at the same position in both histograms.

In order to improve the computational robustness of automatic peak fitting,
after processing all events the histograms are combined binwise: For each bin 
$i$ we
set $h^i = \min(h^i_+,h^i_-)$. This symmetric histogram (see
Fig.~\ref{histexample}) does neither depend on the hit-to-track association
nor on the left-right ambiguity resolution, and even for large offsets a clear
peak can be observed. The fitted peak position is used as an alignment
correction $a^u_k$ and the track reconstruction is repeated iteratively as
long as there is an improvement. 
Beyond a certain alignment accuracy the residual width is dominated by
multiple scattering. Therefore the standard deviation of the residual mean
distribution is used as a control parameter. 
Typically after 20 iterations this quantity is less than 50\,$\mu$m and 
does not improve any further.

\subsubsection{External Degrees of Freedom and Layer $z$ Corrections}

The described algorithm does not fix the alignment constants uniquely.
In general any global transformation that maps straight tracks
into straight tracks will give a valid set of alignment constants:
\begin{displaymath}
\tilde{a}^u_k = a^u_k + [T^x+S^{xz} z_k]\cos\alpha_k
                      - [T^y+S^{yz} z_k]\sin\alpha_k 
\end{displaymath}
Here $T^x$ and $T^y$ are global offsets in $x$ and $y$, $S^{xz}$ and $S^{yz}$
are shearings, i.e.\ offsets that depend linearly on $z$. The stereo angle for
a given GEDE is denoted by ${\alpha_k}$. For each iterative step we
choose the global offsets and shearings such that they minimize the
summed squares of the alignment constants $\sum_k (\tilde{a}^u_k)^2$.

Finally from the alignment constants $\tilde{a}^u_{k_l}$ belonging to layer
$l$ the average $z$ shift $a^z_l$ for this layer can be estimated by
minimizing
$\sum_{k_l} (\langle\tan\theta_{k_l}\rangle a^z_l + \tilde{a}^u_{k_l})^2$.
Here $\langle \tan\theta_{k_l} \rangle$ is the average track slope for the
GEDE plane $k_l$. The error of $a^z_l$ is about a factor
$1/\langle\tan\theta_l\rangle$ larger than the average error of the
$\tilde{a}^u_{k_l}$, $\langle\tan\theta_l\rangle$ being the average track
slope in layer $l$. For the PC chambers one has $\langle\tan\theta_l\rangle
\approx 0.14$, leading to $z$ shift precisions of about 350\,$\mu$m.

\subsubsection{Global Alignment}

The above procedure does not define the position of the OTR within the 
HERA-B coordinate system. To define this global alignment, external tracks 
are necessary. Here the problem appears that the tracking through the 
HERA-B magnetic field depends on the track momentum and the measurement 
of the track momentum depends on the global alignment. To avoid this 
circular argument, special sets of alignment data have been taken with the
magnet switched off. This allows for a straightforward determination of the 
relative position of the OTR.

The data samples used for the global OTR alignment were large enough
to obtain statistical errors of the alignment parameters which were
small compared to the coordinate resolution.
In practice the systematic uncertainties of the parameters were much
bigger than the statistical errors. Their effect on the momentum resolution
will be discussed in Section \ref{sec:momres}.

\section{Detector Performance}
\label{sec:performance}

\subsection{Cell Hit Efficiency}

The cell hit efficiency is the probability to register a signal
in an Outer Tracker drift cell after the passage of a charged 
particle. This is one of the essential quantities which 
characterises the detector performance and which has a strong 
impact on the track and trigger efficiencies.
A charged particle passing a stereo layer should hit at least one 
cell in each single layer and two cells in each double layer.
Counting the layers, the expected number of hits would be 30 
in the PC and 12 in the TC chambers. 
Due to the overlap of cells and modules, the real number 
of hits can be larger.

In order to study the cell hit efficiency, high statistics track samples
have been selected using the following requirements:
\begin{itemize}
\item Events must be reasonably populated, with a total number of hits
   $900 < N_{hits} < 8000$ in the OTR chambers.
\item Tracks must traverse the whole detector, with segments in the Vertex
   Detector (VDS) and in the PC and TC Outer Tracker superlayers.         
\item Tracks must have enough hits per segment, with $N^{VDS} > 6$,
   $N^{TC}+N^{PC} > 20$, and $N^{TC} > 6$.
\item Tracks must have a momentum $p > 5$ GeV/c to keep multiple scattering
   small.
\item Track segments must point to the target, with a vertical coordinate of
   the extrapolated segment at the target position $|y_{target}| < 25$ cm.
\end{itemize}

\begin{figure}[htb] \centering
   \includegraphics[width=\linewidth]{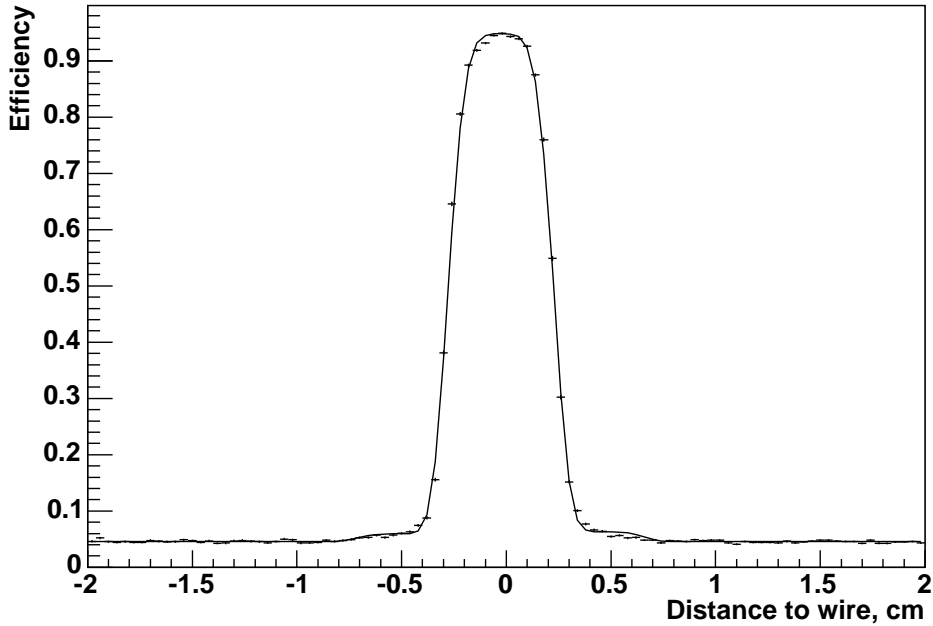}
   \caption{\emph{Example of an efficiency profile for a 5\,mm drift cell.}}
   \label{hist_example}
\end{figure}

A segment is approximated by a straight line connecting the first and last
point of the segment. This line is used to predict the hit positions in all
traversed drift cells.
For each cell two histograms are filled: $N^{found}(x)$ and $N^{expect}(x)$,
i.e.\ the found and expected numbers of hits, respectively, at the distance
$x$ from the segment to the wire. The hit efficiency is calculated as 
\begin{displaymath}
\epsilon_{hit}(x) = \frac{ N^{found}(x) }{ N^{expect}(x) }\,
\end{displaymath}
Fig.~\ref{hist_example} shows an example of such a cell efficiency profile.

To extract the single cell efficiency, the measured efficiency profile is
fitted by a function with six free parameters. The most important ones are the
maximum of the efficiency profile, the smearing of its edges due to cell
geometry and track
parameter errors, the background level which describes the average occupancy,
and the shift of the cell profile due to misalignment.
The method and the fitting procedure are described more 
detailed in \cite{eff_note, eff_note2}.

The resulting average cell efficiency per chamber, i.e.\ the mean value of the
cell efficiency profile maxima, for 5 and 10\,mm cells is summarized in
Table~\ref{tab:XXXXXsoft} for a typical run.
\begin{table}[htb] \centering
   \caption{\emph{Average cell efficiencies (in \%) for 5 and 10\,mm cells in 
       all PC and TC chambers for a standard run.}}
   \vspace*{2mm}
   \label{tab:XXXXXsoft}
   \begin{tabular}{|c|c|c|c|c|}
      \hline\hline
      Superlayer & \multicolumn{2}{c|}{$-x$} & \multicolumn{2}{c|}{$+x$} \\
      \hline
             & 5 mm & 10 mm & 5 mm & 10 mm \\ \hline
      PC1    & 95.8 & 97.5  & 94.9 & 98.2 \\ 
      PC2    & 95.0 & 97.2  & 92.3 & 98.2 \\ 
      PC3    & 96.2 & 98.2  & 96.3 & 98.5 \\ 
      PC4    & 95.0 & 96.8  & 94.2 & 94.7 \\ 
      TC1    & 95.6 & 97.7  & 94.4 & 96.9 \\ 
      TC2    & 92.2 & 95.4  & 89.8 & 93.2 \\
      \hline\hline
   \end{tabular}
\end{table}
The average hit efficiency is about 94\,\% for 5\,mm cells and 97\,\% for
10\,mm cells. This is consistent with the expectation of a higher efficiency
for a larger drift cell at the same gas gain.
The observed lower efficiencies of superlayer TC2 are not 
completely understood.

Hits which are used twofold, first defining the track and then 
estimating the efficiency, can bias the hit efficiency result. 
This was studied by exclusion of a stereo layer from tracking and showed the
bias to be less than 1\,\%.

The First Level Trigger (FLT) of HERA-B utilizes three stereo double layers
in each of the chambers PC1, PC4, TC1, and TC2. In order for a track to be
triggered hit signals must be found in all 12 double layers, with each hit
being the logical OR of two lined-up drift cells. A single cell hit efficiency
$\epsilon_{hit}$ then limits the trigger efficiency to
\begin{displaymath}
   \epsilon_{trig,max} = [1 - (1 - \epsilon_{hit})^2]^{12}.
\end{displaymath}
For $\epsilon_{hit} = 94\,(97)\,\%$ this yields $\epsilon_{trig,max} = 95.8\,
(98.9)\,\%$, which can be compared to the measured efficiency for FLT tracking
in the OTR layers, averaged over the 2002 run, of approximately 60\,\%. The
additional inefficiencies come from limitations of the FLT tracking algorithm
and data transmission problems in the optical links connecting the OTR
front-end electronics to the FLT processor network.

\subsection{Tracking Efficiency}

The tracking efficiency is a basic quantity for most of the physics analyses.
The evaluation method developed for the OTR of HERA-B \cite{pernack} uses a
sample of reference tracks which are reconstructed without using the OTR. A 
reference track consists of a track segment in the VDS together with an
associated RICH ring or ECAL cluster.\footnote{%
  For the purpose of a simplified notation, the following description uses
  VDS-RICH tracks. It likewise holds for VDS-ECAL tracks by replacing every
  occurrence of the term ``RICH'' by ``ECAL''.}
In order to keep the contamination by unphysical tracks
(e.g.\ detector or reconstruction artefacts) 
in the reference sample low,
each reference track must be consistent with originating from a 
$K^0_S \to \pi^+\pi^-$ decay. For all pairwise combinations of reference 
tracks and partner tracks with at least a VDS and an OTR segment the invariant 
mass is calculated, and the fitted number of $K^0_S$ particles in the mass 
spectrum is defined to be the number of reference tracks 
$N^{RICH}_{ref}$.

Then in the vicinity of each reference track an OTR segment matching with the
same VDS segment is searched. If one is found, the track momentum is 
recalculated, and another invariant mass spectrum is generated as described 
above. A fit to this spectrum yields the number of reference tracks confirmed
by the OTR, $N^{OTR}_{ref}$.

\begin{figure}[htb] \centering
   {\unitlength1cm
   \begin{picture}(12.5,7)
      \put( -.5,0.0){\epsfig{file=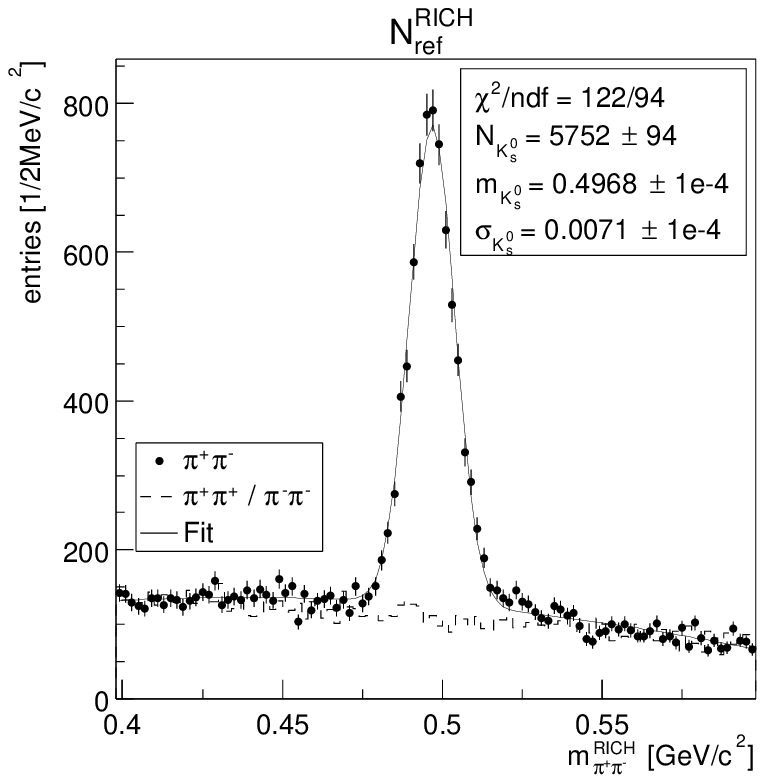,width=6.5cm}} 
      \put(1.2,5.7){\makebox(0,0)[t]{\Large a)}}
      \put(6.0,0.0){\epsfig{file=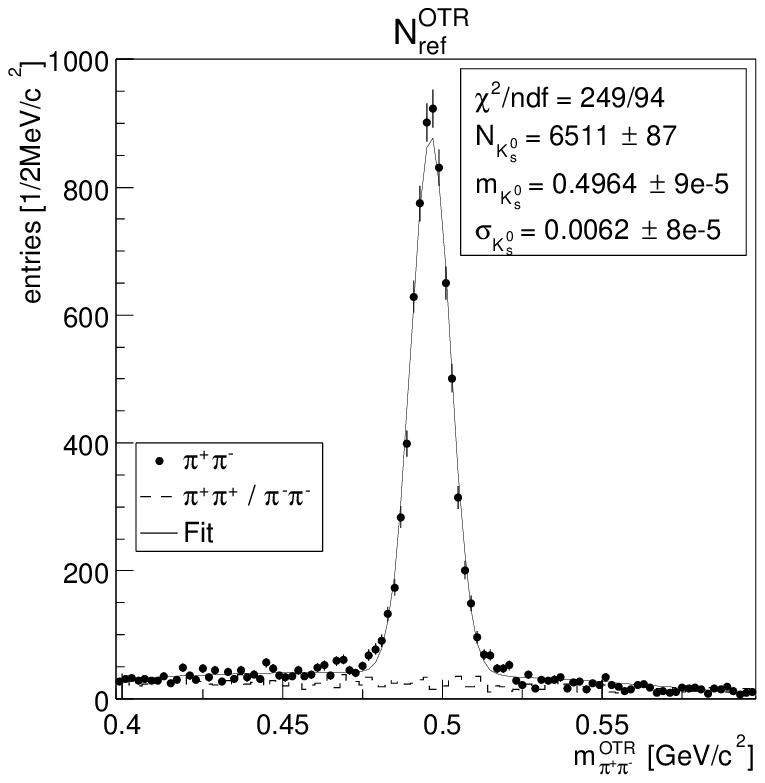,width=6.5cm}} 
      \put(7.7,5.7){\makebox(0,0)[t]{\Large b)}}
   \end{picture}}
   \caption{\emph{Determination of $N^{RICH}_{ref}$ and $N^{OTR}_{ref}$.
      (a) All particle candidates with VDS-RICH reference tracks which
          survive the selection cuts.
      (b) Particle candidates with OTR segments matching the reference tracks.
          The momentum of the reference track is replaced by the momentum of
          the corresponding VDS-OTR track.}} 
   \label{fig:reference}
\end{figure}
Figure \ref{fig:reference} shows an example of corresponding mass spectra and 
fits. The surprising fact that $N^{OTR}_{ref} > N^{RICH}_{ref}$ is due to
the higher probability for wrong matches in the VDS-RICH track sample.
There are $K^0_S$ 
candidates which only show up when the better quality VDS-OTR track is used in
the invariant mass calculation.
\begin{figure}[htb] \centering
   \includegraphics[clip]{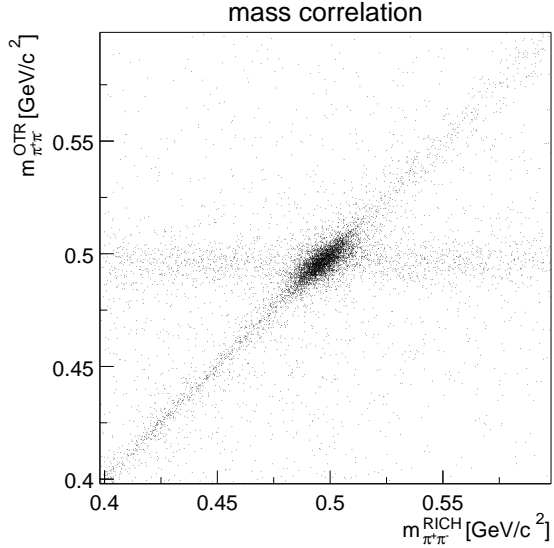}
   \caption{\emph{Correlation between the invariant masses of reference
      particles computed with the momenta from VDS-RICH and VDS-OTR
      tracks, respectively.}}
   \label{fig:mcorrelation}
\end{figure}
This effect can be observed in Fig.~\ref{fig:mcorrelation} where side band 
events in the VDS-RICH sample have the proper invariant mass for the 
corresponding VDS-OTR tracks. The effect of such tracks must be excluded
from the efficiency calculation.

The determination of the number of reference particles $N^{OTR}_{side}$ which
contribute to the $K^0_S$ peak of the $m_{\pi^+\pi^-}^{OTR}$ spectrum, but are
in the side bands of the $m_{\pi^+\pi^-}^{RICH}$ spectrum, is similar to that 
of $N^{OTR}_{ref}$ described above. The only difference is that all 
VDS-RICH reference particle candidates with masses in the range
$m_{K^0_S}^{RICH}\pm 3\sigma_{K^0_S}^{RICH}$ (see Fig.\ \ref{fig:reference}a)
are excluded. A scaling factor $f$ must be introduced to extrapolate the 
number $N^{OTR}_{side}$, which is determined in the side bands, to the full 
mass range. 
Then the tracking efficiency can be calculated as
\begin{displaymath}
   \epsilon = \frac{N^{OTR}_{ref} - f\cdot N^{OTR}_{side}}
                   {N^{RICH}_{ref}}.
\end{displaymath}
The efficiencies evaluated for the OTR in the 2002 running period are 
about 95--97\,\%, as shown in Table~\ref{tab:numbers}.
\begin{table}[htb] \centering
   \caption{\emph{An example of tracking efficiency results for data from the
       2002 running period.}}
   \setlength{\extrarowheight}{2pt}
   \begin{tabular}[htb]{|c|ccccc|}
      \hline\hline
      Mode & $N^{Mode}_{ref}$ & $N^{OTR}_{ref}$ & $N^{OTR}_{side}$ &
       $f$ & $\epsilon$ \\ \hline
      RICH & $7031 \pm 94$ & $8230 \pm 93$ & $1095 \pm 35$ &
             $1.34 \pm 0.02$ & $0.965 \pm 0.020$ \\
      ECAL & $4588 \pm 78$ & $5417 \pm 77$ & $787 \pm 30$ &
             $1.35 \pm 0.02$ & $0.954 \pm 0.026$ \\
      \hline\hline
   \end{tabular}
   \label{tab:numbers}
\end{table}

An important point to be noticed is that the efficiency estimation with this 
method includes both the OTR tracking efficiency and the matching efficiency 
between VDS and OTR segments. This is no disadvantage because the physics 
analyses require the total efficiency anyway. 
In Sec.~\ref{matching} the VDS-OTR matching efficiency is estimated 
to be 98\,\% for pion tracks, which means that the proper
OTR tracking efficiency is even about 2\,\% higher than 
the numbers displayed in Table~\ref{tab:numbers}.

\subsection{Efficiency of Track Matching}
\label{matching}

Track segments are reconstructed independently in the VDS and in the OTR
PC chambers. The vector of track segment parameters at a reference $z$ 
coordinate is
\[ \mathbf{V}^{\rm T} = (x,\; y,\; t_{x},\; t_{y},\; q), \]
where $x$ and $y$ are the transverse coordinates, $t_{x}=p_{x}/p_{z}$
and  $t_{y}=p_{y}/p_{z}$ are the track slopes, and
$q=Q/\left|\vec{p}\right|$ is the ratio of the charge $Q$ and the particle 
momentum $\left|\vec{p}\right|$. 

Matching VDS and PC track segments goes through the following steps: 
\begin{enumerate} 
\item Initially the vectors of segment parameters in the VDS 
   ($\mathbf{V}^i_{VDS}$) and in the PC region ($\mathbf{V}^j_{PC}$) 
   together with their covariance matrices are only partially known, 
   because the component $q$ is not yet defined.
\item A function performing a Kalman filter fit finds all pairs 
   $\mathbf{V}^i_{VDS}$ and $\mathbf{V}^j_{PC}$ which fit to each other. The
   $\chi^2$ of the fit is a measure of the matching quality.
   \label{item:matchfit}
\item For each matched pair the fit yields a track parameter $q$, which is
   common to both segments.
\end{enumerate}

The matching $\chi^2$ in step \ref{item:matchfit} does not follow a $\chi^2$
distribution 
because the track parameter differences from the Kalman filter step
are not Gaussian distributed. 
The reasons are Moli\`{e}re scattering of particles, hit corruption by 
background particles, and failures of pattern recognition due to erroneous 
hit-segment assignment or wrong left-right assignment of drift distances in 
the OTR.
    
\begin{figure}[htb] \centering
\unitlength1cm
\begin{picture}(12.0,6.0)
\put( -0.5,0.0){\epsfig
{file=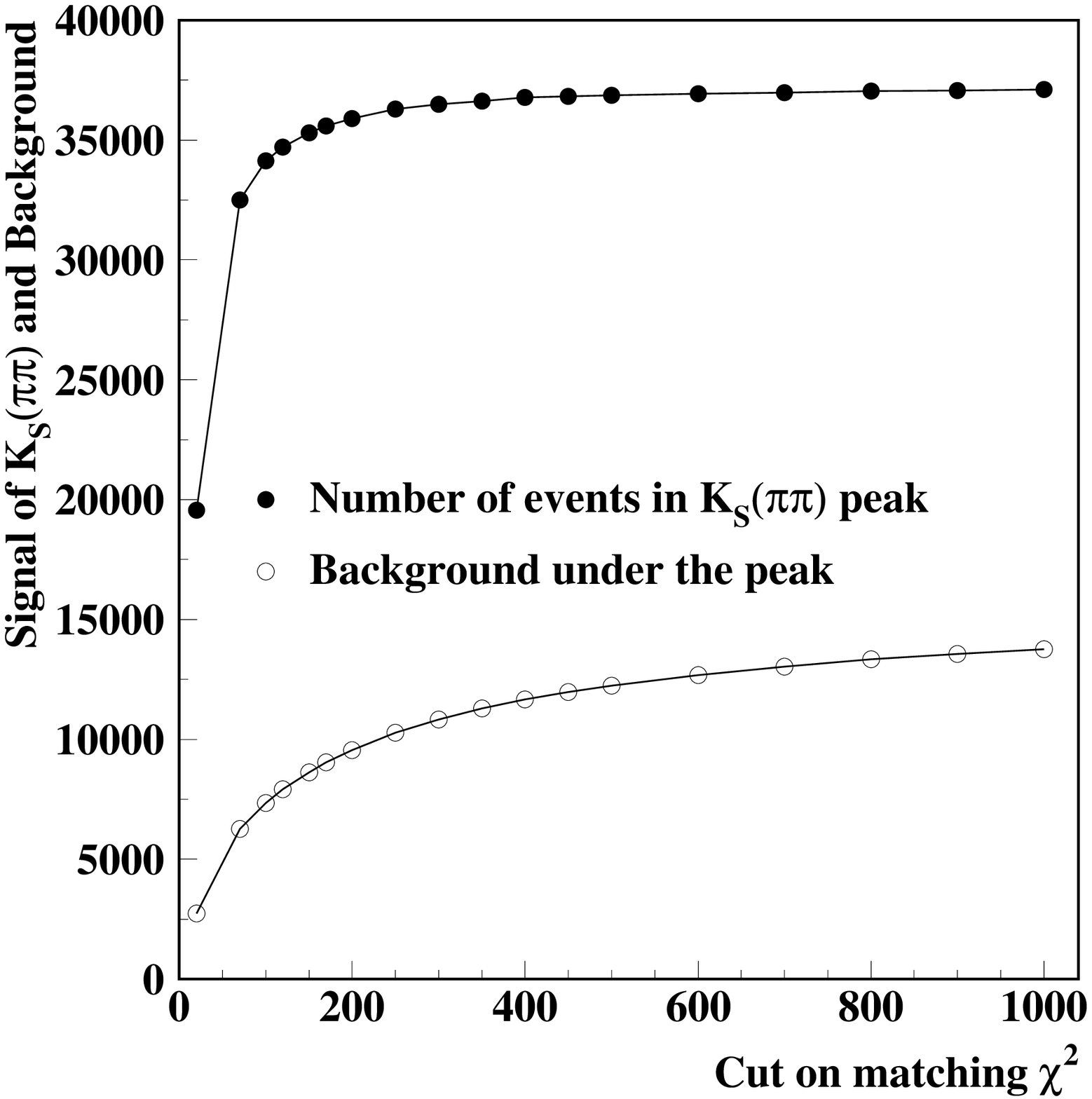,width=6.8cm}}
\put(5.0,2.0){\makebox(0,0)[t]{\Large a)}}
\put(6.1,0.0){\epsfig
{file=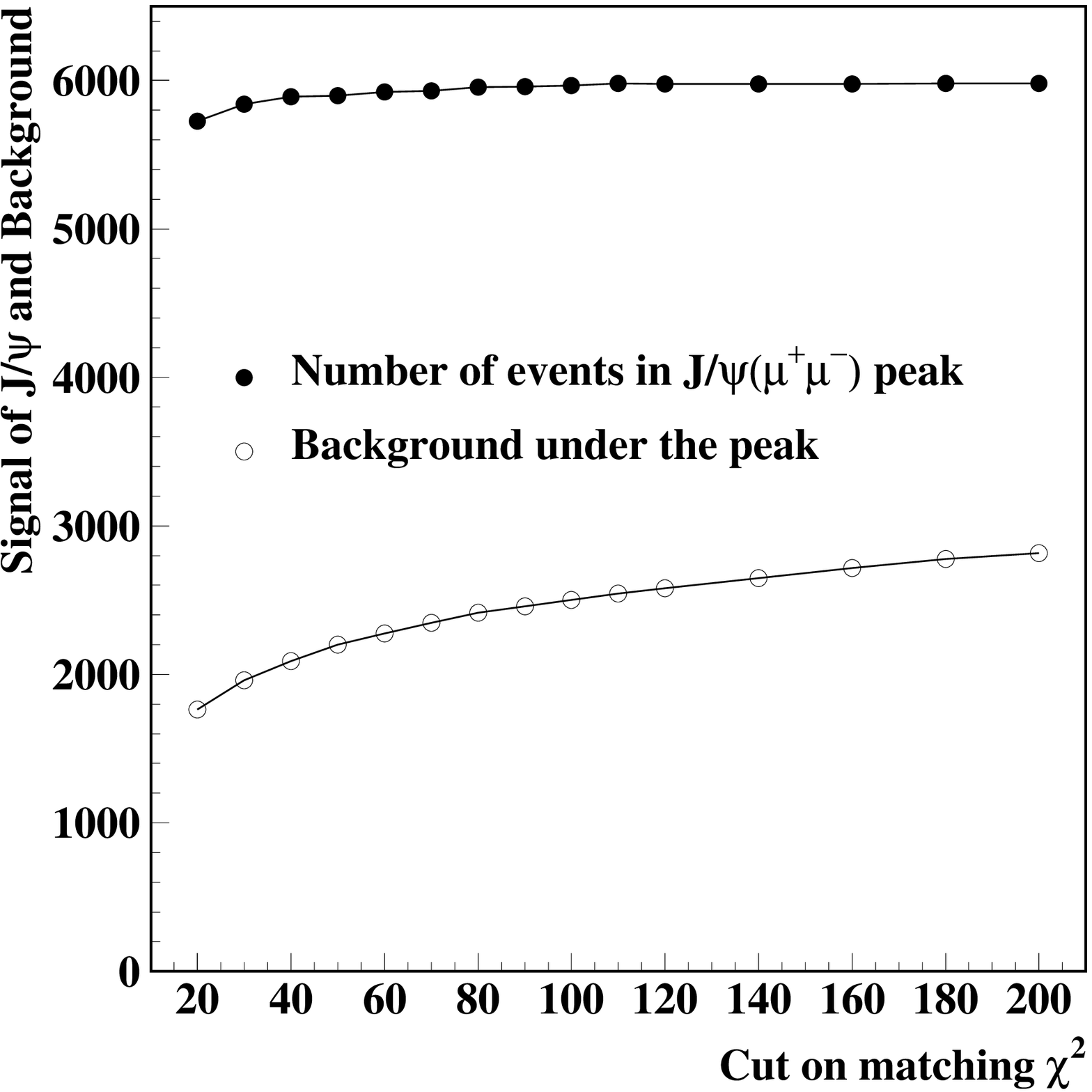,width=6.8cm}}
\put(11.7,2.0){\makebox(0,0)[t]{\Large b)}}
\end{picture}
\caption{The number of (a) $K^0_S\to\pi^+\pi^-$ decays and (b)
   $J/\psi\to\mu^+\mu^-$ decays together with the combinatorial backgrounds
   underneath the signal peaks as functions of the cut on the matching 
   $\chi^2$.}
\label{matchcut}
\end{figure}

In order to find a reasonable cut value for the matching $\chi^2$, the 
reconstruction efficiencies of $K^0_S\to\pi^+\pi^-$ and $J/\psi\to\mu^+\mu^-$
decays were investigated for different cuts on the matching $\chi^2$ 
\cite{matching}.
From the invariant mass spectra of $\pi^+\pi^-$ and $\mu^+\mu^-$ combinations
the number of events in the corresponding peaks and the combinatorial 
background underneath were evaluated. The results are shown in 
Fig.~\ref{matchcut}.

For very soft cuts on the matching $\chi^2$ all proper pairs of segments are
matched and the number of events in the peak reaches a plateau, but also the 
probability to match wrong pairs becomes large and the background increases.
Cutting the matching $\chi^2$ at 200, the default value for data processing,
the number of reconstructed $K^0_S$ decreases by only $(3.3\pm0.1)\,\%$ from 
the plateau value (see Fig.~\ref{matchcut}a).
This results in a matching efficiency of about 98\,\% for a single pion 
track.

The muons from the $J/\psi$ decays are more energetic than the pions
and also have better chances to be reconstructed and matched
because they have already been selected by the trigger.
As can be seen in Fig.~\ref{matchcut}b, the number of reconstructed $J/\psi$
particles already reaches a plateau for matching $\chi^2$ cuts much smaller 
than 200. Hence with the default cut the matching efficiency for energetic
muons is close to 100\,\%.

\subsection{Momentum Resolution}
\label{sec:momres}

The momentum resolution is one of the most important
characteristics of a forward spectrometer \cite{fsres}. It defines not only
the mass resolution, but also the resolutions of
kinematic variables, like the Feynman variable $x_F$ and 
the transverse momentum $p_T$.

\begin{figure}[htb] \centering
\unitlength1cm
\begin{picture}(12.0,5.0)
\put( -0.5,0.0){\epsfig
{file=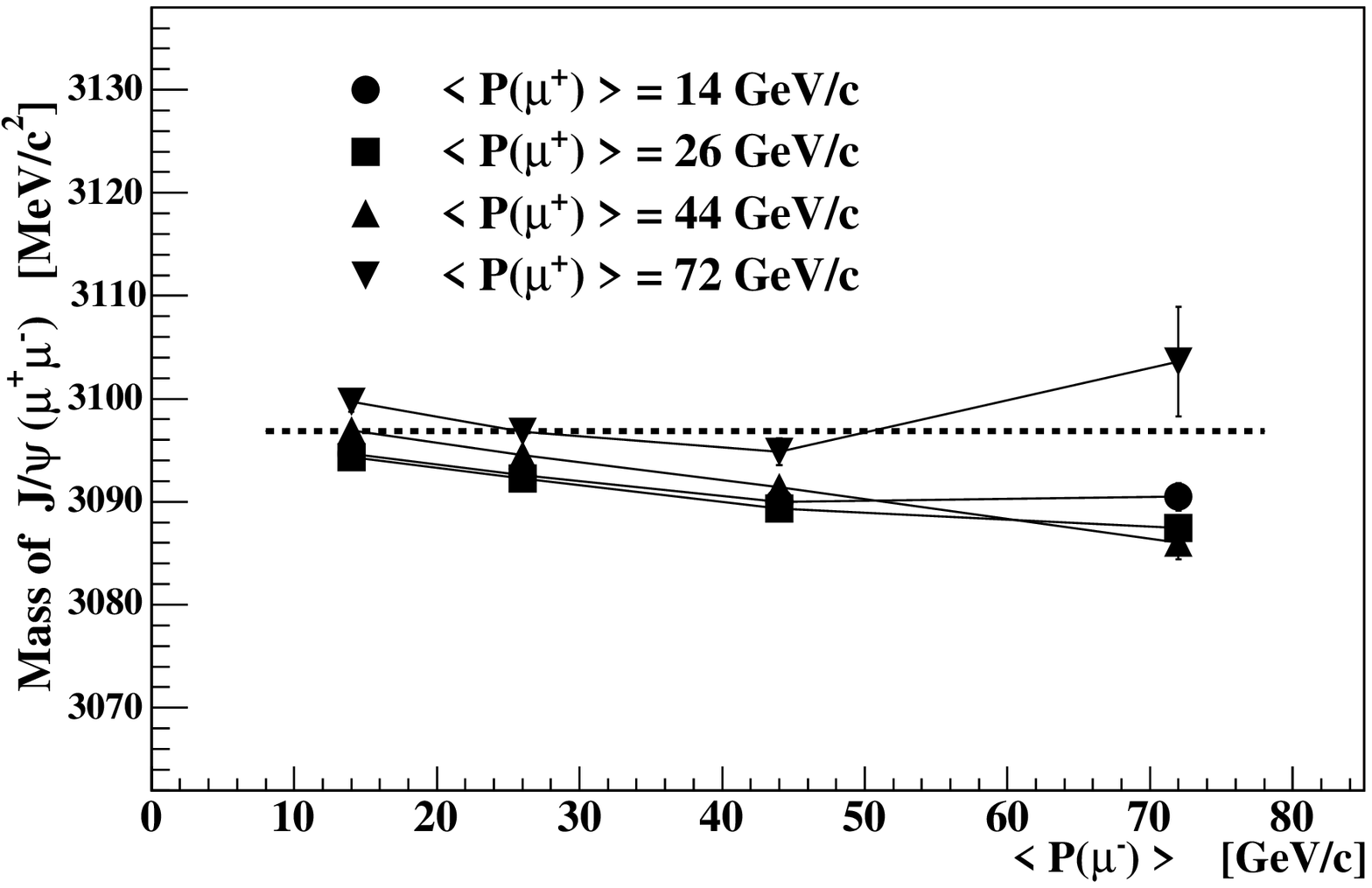,width=7.0cm}} 
\put(5.2,1.5){\makebox(0,0)[t]{\Large a)}}
\put(6.0,0.0){\epsfig
{file=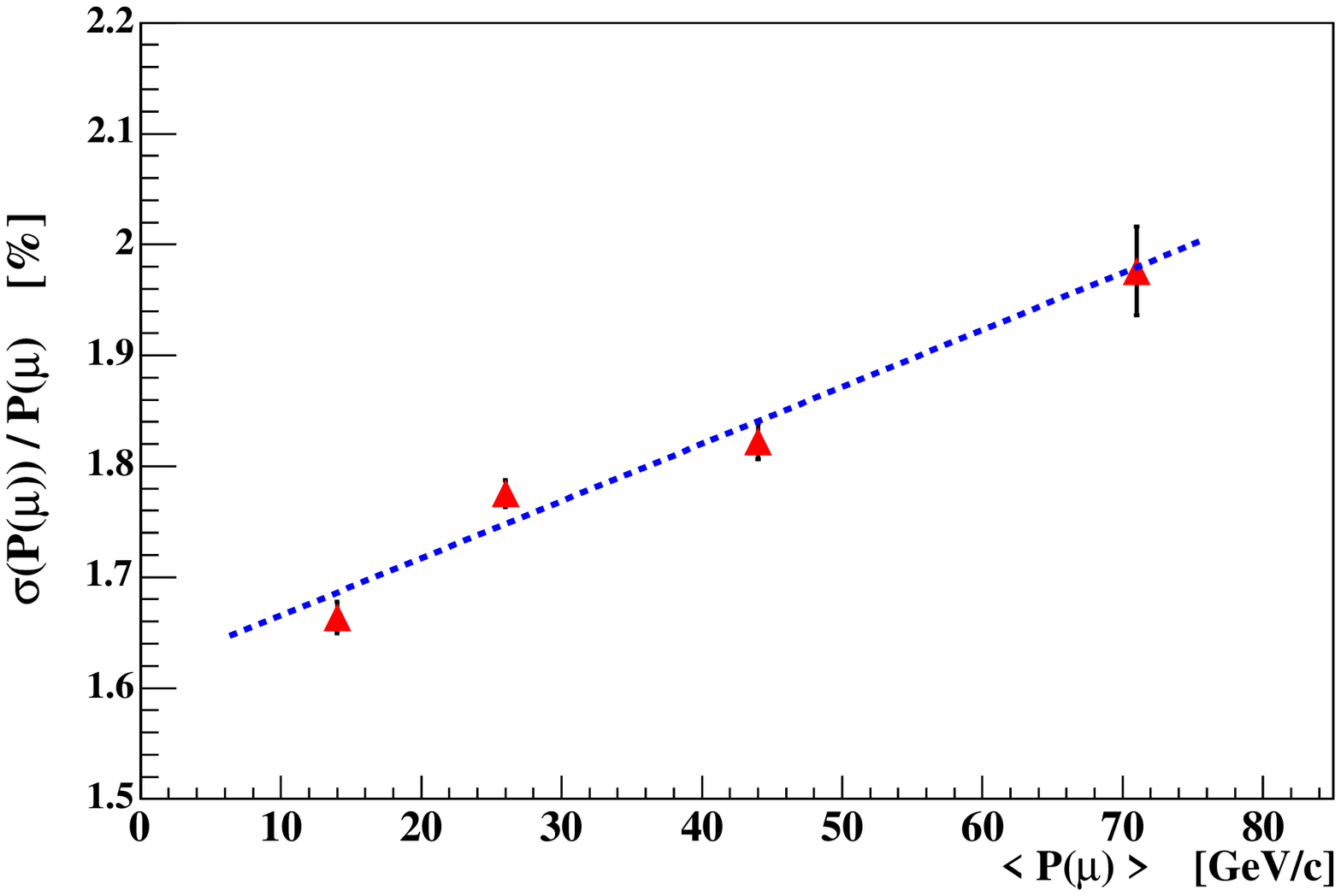,width=7.0cm}} 
\put(11.8,1.5){\makebox(0,0)[t]{\Large b)}}
\end{picture}
\caption{Characteristics of the HERA-B spectrometer
evaluated on a sample of $J/\psi\to\mu^+\mu^-$ decays. 
a) The mean value of the $J/\psi$ mass as a function of the mean $\mu^-$ 
momentum for the given bins of the $\mu^+$ momentum. The dashed line shows the
nominal value of the $J/\psi$ mass.
b) The momentum resolution versus momentum.
The dashed line shows the parametrisation (\ref{parddpp}).}   
\label{refit}
\end{figure}

The performance of the momentum reconstruction procedure was investigated on
a sample of $J/\psi\to\mu^+\mu^-$ decays. This sample was sorted
into a $4 \times 4$ grid, binned in momenta of the $\mu^+$ and the
$\mu^-$. The mean momentum values of each bin were
14, 26, 44, and 72 GeV/c. The mean mass values of the $J/\psi$
in Fig.~\ref{refit}a show good agreement with the nominal
value in all momentum bins, with a maximal relative deviation of 
less than $3\cdot10^{-3}$. 

The mass resolution of the $J/\psi$ can be expressed
by the momentum resolution of muons approximately as
\begin{equation}  
\frac{\sigma(M_{ij})}{M_{ij}} = \frac{1}{2} \left( \frac{\sigma(p_i)}{p_i}
\oplus \frac{\sigma(p_j)}{p_j} \right),
\label{dmtom}         
\end{equation}           
where $i, j$ are the indices of the momentum bins for $\mu^+$ and $\mu^-$,
respectively. The momentum resolution is evaluated by fitting the 
mass resolution in different bins
using the formula (\ref{dmtom}), where the $\sigma(p_i)/p_i$ are kept as
free parameters identical for both muon charges.
The momentum resolution of the HERA-B spectrometer as a function 
of momentum is presented in Fig.~\ref{refit}b. 
In the evaluated momentum range from 10 to 80\,GeV/c the function 
\begin{equation}
\frac{\sigma(p)}{p}\,[\%] = 
(1.61\pm0.02) + (0.0051\pm0.0006) \cdot p \textrm{ [GeV/c]}
\label{parddpp}
\end{equation}
parametrises the momentum resolution rather well \cite{nimmomres}.
Monte Carlo studies in the same momentum region show the same linear
dependence of the momentum resolution on the momentum, but the values
are a factor of about 0.75 lower \cite{mcmomres}. This can be attributed
to remaining detector misalignments which were not simulated.

\section{Conclusions}

In this paper we describe the operation and performance of the HERA-B Outer
Tracker, a 112\,674 channel system of planar drift tube layers. The design and
construction of the detector is described in \cite{otr_design}, the electronics
in \cite{otr_fee} and the aging investigations in \cite{herab_aging}.

The Outer Tracker was run for about 3500 hours in the HERA-B experiment at
target rates of typically 5\,MHz, corresponding to channel occupancies of 
about 2\,\% in the hottest regions. In a first running period the detector 
showed various operational problems
which could be overcome by major repairs and improvements in a longer shutdown.
After that the detector was stably operated at a gas gain of $3\cdot 10^4$
using an Ar/CF$_4$/CO$_2$ (65:30:5) gas mixture. At this working point a good
trigger and track reconstruction efficiency was measured. At the end of the 
HERA-B running no aging effects in the Outer Tracker cells were observed. 
However, the total irradiation stayed far below the originally anticipated 
exposure.

About 95\,\% of all drift cells were well functioning, losses were
mainly due to dead channels while the fraction of noisy channels stayed below
1\,\% in most chambers. Due to the high redundancy of the tracking system this
caused essentially no loss in tracking efficiency which was more than 95\,\% 
for tracks with momenta above 5\,GeV/c.

The hit resolution of the drift cells was 300 to 320\,$\mu$m.
Due to the large lever arm of the tracking system and the high
redundancy the single cell resolution has no large influence on the momentum
resolution which as a function of the momentum in the range from 10 to
80\,GeV/c was determined to be
$$
\frac{\sigma(p)}{p}\,[\%] = (1.61\pm 0.02) + (0.0051 \pm 0.0006)\cdot p
\textrm{ [GeV/c]}
$$

In summary, the performance of the HERA-B Outer Tracker system fullfilled all 
requirements for stable and efficient operation in a hadronic environment,
thus confirming the adequacy of the honeycomb drift tube technology and of the 
front-end readout system. Although the detector has not really been challenged
regarding radiation exposure and occupancies, the performance studies allow to
conclude that 
a suitable tracking concept for a high rate environment was found.

\section*{Acknowledgements}

We thank our colleagues of the HERA-B Collaboration who made in a
common effort the running of the detector possible. The HERA-B
experiment would not have been possible without the enormous effort
and commitment of our technical and administrative staff. It is a
pleasure to thank all of them.

We express our gratitude to the DESY laboratory for the strong support
in setting up and running the HERA-B experiment. We are also indebted
to the DESY accelerator group for the continuous efforts to provide
good beam conditions.

\end{document}